\documentclass[pdftex,twocolumn,epjc3]{svjour3}  

\pdfoutput=1  

\RequirePackage[T1]{fontenc}
\usepackage[utf8]{inputenc}
\RequirePackage{mathptmx}  

\usepackage{amsmath}

\usepackage{graphicx}
\usepackage{float}     
\usepackage{rotating}  

\usepackage{xcolor}
\RequirePackage[colorlinks,citecolor=blue,urlcolor=blue,linkcolor=blue]{hyperref}

\RequirePackage[numbers,sort&compress]{natbib}

\usepackage{orcidlink}
\usepackage{comment}
\RequirePackage{flushend}  
\smartqed  
\journalname{Eur. Phys. J. C}
\begin{document}

\title{Constraining Viscous-Fluid Models in $f(Q)$ Gravity with Data}

\author{
    Shambel Sahlu \orcidlink{0009-0005-6501-8433}\thanksref{addr1,addr2,e1} \and
    Renier T. Hough \orcidlink{0000-0003-0316-8274}\thanksref{addr1,e2} \and
    Amare Abebe \orcidlink{0000-0001-5475-2919}\thanksref{addr1,addr3,e3} \and
    \'{A}lvaro de la Cruz-Dombriz \orcidlink{https://orcid.org/0000-0002-7072-9396}\thanksref{addr4,addr5,e4}
}

\thankstext{e1}{e-mail: \texttt{shambel.sahlu@nithecs.ac.za} (corresponding author)}
\thankstext{e2}{e-mail: \texttt{25026097@mynwu.ac.za}}
\thankstext{e3}{e-mail: \texttt{amare.abebe@nithecs.ac.za}}
\thankstext{e4}{e-mail: \texttt{alvaro.dombriz@usal.es}}

\institute{
    Centre for Space Research, North-West University, Potchefstroom 2531, South Africa \label{addr1} \and
    Department of Physics, Wolkite University, Wolkite, Ethiopia \label{addr2} \and
    National Institute for Theoretical and Computational Sciences (NITheCS), Potchefstroom 2520, South Africa \label{addr3} \and
    Cosmology and Gravity Group, Department of Mathematics and Applied Mathematics, University of Cape Town, 7700 Rondebosch, South Africa \label{addr4} \and
    Departamento de F\'isica Fundamental, Universidad de Salamanca, 37008 Salamanca, Spain \label{addr5}
}

\date{Received: date / Accepted: date}

\maketitle

\begin{abstract}

We investigate the impact of bulk viscosity on the accelerating expansion and large-scale structure formation of a Universe in which the underlying gravitational interaction is described by $f(Q)$ gravity. Various paradigmatic choices of the $f(Q)$ gravity theory, including power-law, exponential, and logarithmic models, are considered. To test the cosmological viability of these $f(Q)$ gravity models, we use {the Baryon Acoustic Oscillations ($BAO$)  measurements from the Dark Energy Spectroscopic Instrument (DESI) Survey, cosmic chronometers ($CC$)  from Hubble measurements, the SNIa distance moduli measurements from the \textit{PantheonP + SH0ES}, growth rate ($f$-data), and redshift-space distortions ($f\sigma_8$) datasets, the latter two once the linear cosmological perturbations, growth rate \(f(z)\), and redshift-space distortion \(f\sigma_8(z)\) are studied. 
Thus, we perform the combined analyses for: \textit{PantheonP + SH0ES}, \textit{PantheonP + SH0ES} + \textit{f}, and  \textit{PantheonP + SH0ES} + $f\sigma_8$}. We compute the best-fit values \(\Omega_m\),  $H_0\,\mathrm{(km/s/Mpc)}$,  \(r_d\), \(M_{abs}\), \(\gamma\), \(\sigma_8\), \(n\), \(p\) and \(\Gamma\) including the bulk viscosity coefficient \(\zeta\). Through a detailed statistical analysis, based on the Akaike Information Criterion (AIC) and Bayesian / Schwartz Information Criterion (BIC), a statistical comparison of the $f(Q)$ gravity models with $\Lambda$CDM is made. 
Among the three $f(Q)$ models, only the non-viscous $f(Q)$ power-law model yields robust parameter estimates and substantial observational support without any outright rejections. In contrast, both exponential and logarithmic $f(Q)$ models (with or without bulk viscosity) are rejected by multiple model selection criteria. Moreover, adding bulk viscosity consistently increases $\Delta$AIC and $\Delta$BIC  values, indicating that its inclusion is not statistically justified.

\end{abstract}

\section{Introduction}\label{sec:intro}
Although $\Lambda$CDM is the current leading cosmological model in explaining most of the observed phenomena in the Universe, its limitations highlight the need for alternative theories. \(f(Q)\) gravity is one of the latest entrants in an already crowded field of modified theoretical extensions of Einstein's General Relativity (GR) \cite{ferraris1982variational}, {{most of them} designed to address different aspects of cosmological problems such as the puzzle of the accelerated expansion of the Universe, without resorting to the inclusion of a dark energy component. Unlike GR, which employs the Levi-Civita connection to characterize spacetime curvature within the Riemannian geometric framework, $f(Q)$ gravity is based on the non-metricity tensor $Q$ \cite{heisenberg2024review}, rather than curvature or torsion.  Indeed, both Teleparallel Equivalent of GR (TEGR), whose gravitational Lagrangian is proportional to $T$,  and Symmetric Teleparallel Equivalent of GR (STEGR), whose gravitational Lagrangian is proportional to $Q$, can
be shown to be equivalent theories, and equivalent to General Relativity, since they differ by a total derivative should the choice of
connection be the adequate one. We refer the interested reader to seminal references such as \cite{beltran2019geometrical}. 
Nonetheless, once these invariants, $T$ and $Q$, are dressed by an arbitrarily general - not linear - function $f$ of the respective argument, the
equivalence between $f(Q)$ and $f(T)$ vanishes. The equations of motion are different and, in general, they host a different number of
degrees of freedom propagating, so with that, you can show that they are not equivalent. This tensor is a fundamental measure of how the length of vectors is altered under parallel transport in a non-Riemannian manifold, which departs from the usual curvature-based description of spacetime. By exploring the implications of this non-metricity formulation, $f(Q)$ gravity seeks to offer new theoretical insights and potentially more accurate descriptions of the Universe's acceleration \cite{koussour2022cosmic}, {the evolution of large-scale structure dynamics \cite{khyllep2023cosmology, Khyllep}}, early inflation \cite{capozziello2022slow}, 
and the behavior of compact astrophysical objects like black holes and neutron stars \cite{d2022black,das2024neutron,das2024spherically}. Modifications in $f(Q)$ gravity might also influence the propagation of gravitational waves \cite{capozziello2024gravitational}, offering new ways to test the theory through observational data. 
These inquiries predominantly hinge on theoretical predictions of the cosmological expansion history, confirmed with empirical data such as Supernovae Type Ia ({SNIa}), Baryon Acoustic Oscillations (BAO), and the Cosmic Microwave Background (CMB) shift factor, and provide critical insights into the expansion history, underscoring the keen implications of $f(Q)$ gravity \cite{shekh2023observational}. 
%

On the other hand, in the purely Einsteinian (GR) context, viscosity effects in cosmic fluids have been widely studied to address key aspects of cosmic evolution \cite{barrow1986deflationary,gron1990viscous,  maartens1995dissipative}. In fact, bulk viscosity is the only viscous influence that can change the
background dynamics in a homogeneous and isotropic universe endowed
with the usual GR geometric content. By incorporating bulk viscosity, the effective pressure $p\propto \zeta_1 H$ helps explain phenomena like accelerated expansion and entropy production without a cosmological constant \cite{gron1990viscous}. 

Furthermore, the effect of introducing imperfections to the fluid, such as bulk viscosity, has recently gained traction \cite{chimento1997cosmological, maartens1995dissipative, maartens1997density, colistete2007bulk, wilson2007bulk, acquaviva2016dark}. During early cosmic epochs, bulk viscosity influences the behavior of the cosmic fluid, particularly during phases such as reheating and phase transitions \cite{boyanovsky2006phase}. In inflationary scenarios, it contributes to deviations from the behavior of perfect fluids, facilitating slow-roll conditions and influencing primordial perturbations. Bulk viscosity also imprints on the CMB by dampening acoustic oscillations, therefore altering the observed anisotropies and the power spectrum \cite{gagnon2011dark}. Furthermore, it is instrumental in models that explain the Universe's late-time accelerated expansion without invoking dark energy, where a decaying bulk viscous pressure modifies the Friedmann equations \cite{wilson2007bulk}.
Furthermore, bulk viscosity may also play a role in  gravitational collapse \cite{blas2015large} and structure formation, thus affecting density perturbations \cite{graves1999bulk,li2014viscous,acquaviva2016dark}. Hence, integrating the presence of bulk viscosity into geometrically-motivated cosmological models is essential for a comprehensive understanding of the cosmological dynamics and large-scale structure of the Universe \cite{blas2015large}.

{As per the above,} the combination of $f(Q)$ framework fueled by viscous fluids 
has already been a topic of some interest so far. For instance, in \cite{rana2024phase} authors broadly addressed the theoretical aspects of the expansion-history only phase space analysis for cosmological models that involve viscous fluids within the \( f(Q) \) gravity framework. It does so by incorporating bulk viscous pressure into the matter fluid through an effective pressure mechanism.
Additionally, in \cite{gadbail2021viscous}, the impact of the bulk viscous parameter was tested within the dynamics of the Universe in a Weyl-type $f(Q, \mathcal{T})$ gravity, where $Q$ is the non-metricity and $\mathcal{T}$ is the trace of the matter energy-momentum tensor. 
In summary, for the first time in literature in this study we aim {to examine the effects of bulk viscosity on both the late-time acceleration of the Universe and the matter density fluctuations within the framework of $f(Q)$ gravity, exemplified with three paradigmatic models. Through MCMC simulations, for the former, we use cosmic measurements to determine the best-fit cosmological parameters. Then, for the latter, we further explore the density fluctuations evolution equations using the 1+3 covariant formalism and their ability to fit large-scale structure measurements.}

The  article is structured as follows: Section \ref{theory} provides a comprehensive review of the theoretical framework of $f(Q)$ gravity cosmology and introduces the modified Friedmann equation under the assumption that the bulk viscosity influences the effective pressure. Therein, in Section \ref{models} we present three viable $f(Q)$ models and highlight the key expressions necessary for parameter constraints based on cosmic expansion measurements. Then, in Section \ref{perturbations} we provide the key equations for the growth of scalar perturbations in this context. By resorting to them, the growth rate and the redshift space distortion for the three $f(Q)$ gravity models under study ban be extracted. The article proceeds in  Section \ref{datameth} by describing the methodology and data used in this work. Subsequently, in Section \ref{resultdiscussion}, the results of this work are presented and discussed. In this section the cosmological parameters and the statistical analysis are presented using the $f(Q)$ gravity models for both cases: (i) in the presence of the bulk viscosity effect and (ii) in its absence.  Finally, in Section \ref{conculusions}, we present our conclusions. Throughout this communication $8\pi G=c=1$ units are used.

\section{Theory}\label{theory}
The  $f(Q)$ gravity theory is developed using the metric-affine formalism, where the gravitational field is described by both the metric tensor \(g_{\mu\nu}\) and a connection \(\Gamma^\lambda_{\mu\nu}\). The action is typically written as \cite{jimenez2018coincident}
\begin{equation}\label{eq1}
	S=\int\sqrt{-g} \left(\frac{1}{2}f(Q)+\mathcal{L}_m\right) {\rm d}^4x\;, 
\end{equation}
where
\( g \) is the determinant of the metric tensor \( g_{\mu\nu} \), \( Q \) is the non-metricity scalar, which describes the gravitational interaction, \( f(Q) \) is a general function of the non-metricity scalar, \( \mathcal{L}_m \) is the matter Lagrangian.
%
As customary, the energy-momentum tensor corresponds to $T_{\mu\nu} = -\frac{2}{\sqrt{-g}}\frac{\delta(\sqrt{-g}\mathcal{L}_m)}{\delta g^{\mu\nu}}$, and the non-metricity tensor is the covariant derivative of the metric tensor
\begin{eqnarray}
Q_{\gamma\mu\nu} \equiv \nabla_\gamma g_{\mu\nu}\;, \qquad Q_\gamma = Q_{\gamma}~^{\mu}~_{\mu} \qquad \tilde{Q}_\gamma = Q^{\mu}~_{\gamma\mu}\,,
\end{eqnarray}
while the superpotential term $P^\gamma_{\mu\nu}$ is given by
\begin{eqnarray}
&&P^\gamma_{\mu\nu} = \frac{1}{4}\left( - Q^\gamma~_{\mu\nu}+2Q_{(\mu}~^\gamma~_{\nu )} - Q^\gamma g_{\mu\nu} - \tilde{Q}^\gamma g_{\mu\nu} - \delta ^\gamma_{(\gamma}Q_{\nu)}\right)\,,\nonumber\\
&&
\end{eqnarray}
and the trace of nonmetricity tensor, $Q \equiv -Q_{\mu\nu\gamma}P^{\mu\nu\gamma}.$
For our case, we consider the gravitational action with minimal coupling between the matter and the connection, as described in Eq.\ref{eq1}.} In the following, we assume a spatially flat Friedmann-Lema\^{i}tre-Robertson-Walker (FLRW) metric
\begin{eqnarray}
    {\rm d}s^2 = -{\rm d}t^2 +a^2(t)\delta_{\alpha\beta}{\rm d}x^\alpha {\rm d}x^\beta\;,
\end{eqnarray}
where $a(t)$ is the cosmological scale factor.\footnote{$\delta^{\alpha}{}_{\beta} =1$, when $\alpha = \beta$, zero otherwise.} The corresponding modified Friedmann equation once the gravitational Lagrangian has been split into $f(Q) = Q +F(Q)$ becomes
\begin{eqnarray}\label{friedman}
	3H^2= \rho_m + \rho_r+\rho_{de} \;,
\end{eqnarray}
where $\rho_m$, $\rho_r$ and $\rho_{de}$, is the energy density of matter, radiation\footnote{From now onwards, we shall assume the radiation component is negligible since our focus would be in the late universe}, and dark energy, respectively, the latter produced by the contribution of $f(Q)$ gravity. The total thermodynamic quantities are defined as
\begin{eqnarray}\label{densitypressure}
	 \rho_{tot} = \rho_m +\rho_{de}\;,~~\mbox{and } ~~ p_{tot} = \bar{p} +p_{de}\;.
\end{eqnarray}
Therein, the contributions from the $f(Q)$ gravity correspond to
\begin{eqnarray}\label{pressure}
	&&\rho_{de}=\frac{F}{2}-QF',\label{55}~~p_{de} = -\rho_{de}+2\dot{H} (2QF''+F')\;. \label{555}
\end{eqnarray}
{where prime and dot denote derivative with respect to $Q$ and cosmic time $t$ respectively. On the other hand,} 
the stress-energy tensor in the presence of a viscous fluid satisfies
\begin{eqnarray}\label{EMT}
    T_{\mu\nu} = (\rho +\bar{p})u_\mu u_\nu +\bar{p} g_{\mu\nu}\;, 
\end{eqnarray}
where $\bar{p}$ is effective pressure of the matter and $u_\mu$ is the four-velocity vector of the fluid. Then the effective pressure for matter fluid reads $\bar{p} = p_m +\zeta_1(\rho)H$. It is assumed that the bulk viscosity coefficient is related to the energy density of the matter fluid as \( \zeta_(\rho) = \zeta \rho^\delta \) \cite{barrow1986deflationary,gron1990viscous,  maartens1995dissipative}, where \( \delta \geq 0 \). We have simplified by assuming the linear form of the bulk viscosity coefficient, setting \( \delta = 0 \). The conservation equations read as \cite{rana2024phase,barrow1986deflationary,gron1990viscous,  maartens1995dissipative}\footnote{Since the contribution of radiation to the late-time cosmological expansion history is so minimal, we have safely neglected such a contribution in our analysis.}
\begin{eqnarray}
    && \dot{\rho}_{m}+3H\left(\rho_m-\zeta\rho_m\right) = 0\;, \label{energymatter}\\&&
    \dot{\rho}_{de}+3H\left(\rho_{de}+p_{de}\right) = 0\;.
\end{eqnarray}
From Eq. \eqref{energymatter}, the solution for the energy density of the matter fluid is given by $\rho_m = \rho_{m,0}(1+z)^{3(1-\zeta)} = 3H^2_0\Omega_m (1+z)^{3(1-\zeta)}$, where $H_0\,\mathrm{(km/s/Mpc)}$ is the present-day value of the Hubble parameter, $\Omega_m$ is the present-day value of the fractional energy density and $z$ is the cosmological redshift. For the case of $\zeta= 0$, the energy-momentum tensor Eq. \(\eqref{EMT}\) recovers the perfect fluid expression whereas, as a consequence of the usual conservation equation in  
\eqref{energymatter} the matter fluid recovers its usual redshift dependence.

\subsection{Viable $f(Q)$ models}\label{models}
In the following sections we shall consider three viable $f(Q)$ models - labeled $f_{1\rm{CDM}}$, $f_{2\rm{CDM}}$ and $f_{3\rm{CDM}}$ -
in order to explore the expansion history and the growth of large-scale structures, both with and without accounting for the influence of bulk viscosity. Under certain choice of their parameters, such models can recover the Usual Einstein-Hilbert action - in the $Q$ picture - endowed with or without a cosmological constant. They are as follow

\subsubsection{$f_{1\rm{CDM}}$}
This power-law model \cite{Khyllep,rana2024phase} is given by 
\begin{eqnarray}\label{model1}
        F(Q) = \alpha Q^n\;,
    \end{eqnarray}
where $\alpha$ and $n$ are constants. For the case of $n = 0$, the $f(Q)$ gravity model reduces to $\Lambda$CDM. Similarly, when $n = 1/2$, the above modified Friedmann Eq. \eqref{friedman} simplifies to GR. From Eq. \eqref{friedman}, in this scenario we obtain 
\begin{equation}\label{modelalpha}
\alpha = ({1-\Omega_m)}/{(6H_0^2)^{n-1}(2n-1)}.
\end{equation}
Straightforwardly, the normalized Hubble parameter $E$ from \eqref{friedman} and \eqref{model1} yields
    \begin{eqnarray}\label{power}
    E^{2} \equiv \frac{H^2(z)}{H_0^2} = {\Omega_m}(1+z)^{3(1-\zeta)}+(1-\Omega_m)E^{2n}\,.
\end{eqnarray}
For the case of $\zeta= 0$, the normalized Hubble parameter reduces to \cite{Khyllep} \begin{eqnarray}
    E^2  = {\Omega_m}(1+z)^3+(1-\Omega_m)E^{2n}\,.
\end{eqnarray}

\subsubsection{$f_{2\rm{CDM}}$}
The exponential model \cite{gadbail2024modified} reads as 
\begin{eqnarray}\label{model2}
        F(Q) = \beta Q_0\left(1-{\rm e}^{-p\sqrt{\frac{Q}{Q_0}}}\right)\;,
    \end{eqnarray}
where $p$ is the constant and $Q_0$ is today's value of the non-metricity tensor. For the case of $p = 0$, the model reduces to GR. The parameter $\beta$ is obtained from Eq. \eqref{friedman} yielding
\begin{eqnarray}
\beta  = ({1-\Omega_m})({1-(1+p){\rm e}^{-p}})^{-1}, 
\end{eqnarray}
while the modified normalized Hubble parameter for this exponential model in the presence of bulk viscosity becomes 
    \begin{eqnarray}
    \label{expo}
       && E^2 = \Omega_m(1+z)^{3(1-\zeta)} + (1-\Omega_m)\frac{\left[1-(1+p) E {\rm e}^{-p E}\right]}{1-(1+p)e^{-p}}\;.
       \nonumber\\&&
       \label{expo}
    \end{eqnarray}

\subsubsection{$f_{3\rm{CDM}}$}
For the case of the logarithm model \cite{gadbail2024modified}, it can be expressed as
\begin{eqnarray}
        F(Q) = \epsilon \ln{\left(\Gamma \frac{Q}{Q_0}\right)}\;,
    \end{eqnarray}
where $\epsilon$ and $\Gamma > 0$ are constants and the parameter $\epsilon$ is obtained from Eq. \eqref{friedman} as $\epsilon  = ({1-\Omega_m})Q_0/(2-\ln(\Gamma))$. The modified normalized Hubble parameter for the exponential model in the presence of bulk viscosity becomes 
    \begin{eqnarray}\label{log}
    && E^2 = \Omega_m(1+z)^{3(1-\zeta)} + \frac{1-\Omega_m}{2- \ln(\Gamma)}
    \left( 2-\ln\left(\Gamma E\right)\right)\,.
    \end{eqnarray}
 
\subsection{Cosmological perturbations in $f(Q)$ gravity models}\label{perturbations}
To study the structure growth in a viscous fluid with these three \(f(Q)\) gravity models, this section explores the formulation of the density contrast \(\delta(z)\) using the \(1+3\) covariant formalism \cite{dunsby1991gauge,dunsby1992cosmological,ellis1989covariant,sahlu2023confronting,sahlu2020scalar, abebe2013large,abebe2012covariant}. Later, we will constrain the cosmological parameters \(\{\Omega_m, \sigma_8\}\), \(n, p, \Gamma\), and \(\zeta\), and present a detailed statistical analysis using large-scale structure data, specifically {f} and {f}$\sigma_8$. As presented in the work \cite{castaneda2016some}, the Raychaudhuri equation for a perfect fluid is 
 \begin{eqnarray}\label{raychuadhuri}
   \dot{\theta}=-\frac{1}{3}\theta^2-\frac{1}{2}(\rho_{tot}+3p_{tot})+\nabla^a\dot{u}_a\;, 
 \end{eqnarray}
where $u_a$ is the four-vector velocity of the matter fluid. Substituting Eqs. \eqref{densitypressure} and \eqref{pressure} into Eq. \eqref{raychuadhuri}, the Raychaudhuri equation for $f(Q)$ gravity is given as \cite{sahlu2025structure}
\begin{equation}
	\dot{\theta}=-\frac{1}{3}\theta^2-\frac{1}{2}(\rho+3\bar{p})+\rho_{de}-\frac{\theta\dot{Q}}{2Q}\left(2QF''+F'\right) +\nabla^a\dot{u}_a \;,
\end{equation}
where $\tilde{\nabla}_a$ represents the covariant spatial gradient and $\theta \equiv 3H$. For the cosmological background, we consider a homogeneous and isotropic expanding Universe, taking into account the spatial gradients of gauge-invariant variables such as
\begin{eqnarray}
	&&D^m_a=\frac{a}{\rho_m}\tilde{\nabla}_a\rho_m\;,\label{35} \qquad 
	Z_a=a\tilde{\nabla}_a\theta\;,\label{36}
\end{eqnarray}
where $D^m_a$ represents the energy density and \(Z_a\) stands for the volume expansion of the fluid as presented in \cite{dunsby1992cosmological, abebe2012covariant}. These variables are essential for deriving the evolution equations for density contrasts. The new terms $\mathcal{W}_a$ and $\mathcal{L}_a$ are introduced in \cite{sahlu2025structure}, for the spatial gradients of gauge-invariant quantities to characterize the non-metricity density and momentum fluctuations, respectively based on the non-metricity for an arbitrary $f(Q)$ gravity model, and is expressed as 
\begin{eqnarray}
	&&\mathcal{W}_a=a\tilde{\nabla}_aQ,\label{ggg} \qquad \mathcal{L}_a=a\tilde{\nabla}_a\dot{Q}\;.\label{oooo}
\end{eqnarray}
By admitting the definition Eqs. \eqref{36} - \eqref{oooo}, we have implemented the scalar decomposition method to find the scalar perturbation equations for $f(Q)$ gravity, which are responsible for the formation of large-scale structures \cite{ellis1989covariant,dunsby1991gauge, dunsby1992cosmological, abebe2012covariant,sahlu2020scalar, sami2021covariant, Ntahompagaze, sahlu2025structure}. To extract any scalar variable $Y$ from the first-order evolution equations, we perform the usual decomposition yielding
	\begin{equation}
		a\nabla_aY_b=Y_{ab}=\frac{1}{3}h_{ab}Y+\Sigma_{ab}^Y+Y_{[ab]}\;.
	\end{equation}
Here $Y=a\nabla_a Y^a$, whereas $\Sigma^Y_{ab}=Y_{(ab)}-\frac{1}{3}h_{ab}Y$ and $Y_{[ab]}$ 
represent the shear (distortion) and vorticity (rotation) of the density gradient field, respectively. Then, we define the following scalar quantities as \cite{abebe2012covariant,sahlu2025structure}
\begin{equation}
    \mathcal{\delta}_m=a\tilde{\nabla}^aD^m_a,~Z=a\tilde{\nabla}^aZ_a,~\mathcal{W}=a\tilde{\nabla}^a\mathcal{W}_a,~\mathcal{L}=a\tilde{\nabla}^a\mathcal{L}_a.
\end{equation}
After employing, the scalar decomposition, we also conducted the harmonic decomposition method as detailed in \cite{abebe2012covariant, ntahompagaze2018study, sami2021covariant} to determine the eigenfunctions with the corresponding wave number $\tilde{\nabla}^2 \equiv -{k^2}/{a^2}$, where the wave number $k = \frac{2\pi a}{\lambda}$ \cite{dunsby1992cosmological} and $\lambda$ represents the wavelength. This approach is used to solve harmonic oscillator differential equations in $f(Q)$ gravity. The harmonic decomposition technique is applied to the first-order linear cosmological perturbation equations of scalar variables to derive the eigenfunctions and wave numbers \cite{sahlu2020scalar}.
\\
\\
For any second-order functions $X$ and $Y$, the harmonic oscillator equation is given as
	\begin{equation}
		\ddot{X}=A\dot{X}+BX-C(Y,\dot{Y} ),
	\end{equation}	
	where the frictional force, restoring force, and source force are expressed by $A$, $B$, and $C$, respectively, and the separation of variables takes the form
		$X=\sum_{k}X^k(t)Q^k(x), \hspace{0.1cm}{\rm and}  \hspace{0.1cm}	Y=\sum_{k}Y^k(t)Q^k(x),$ where $k$ is the wave number and $Q^k(x)$ is the eigenfunction of the covariantly defined Laplace-Beltrami operator in (almost) FLRW space-times, $\tilde\nabla^2Q^k(x)=-\frac{k^2}{a^2}Q^k(x).$
The work in \cite{sahlu2025structure} comprehensively analyses the scalar and harmonic decomposition techniques within the framework of $f(Q)$ gravity models, as presented in
{\small
  \begin{eqnarray}
		&&\ddot{\delta}_m^k= -\Bigg[ \frac{2\theta}{3}+\frac{\dot{Q}F'}{2Q}+\dot{Q}F''-w\theta\Bigg]\dot{\delta}_m^k+\Bigg[w F+\theta\dot{Q}F''w  \nonumber\\&& -\frac{\theta\dot{Q}F'w}{2Q}+\frac{(1+3w)\rho_m}{2}(1-w) \label{70011} -\frac{k^2}{a^2} w\Bigg]\delta_m^k+\Bigg[ \frac{1}{2}F'   +QF'' \\&& -\theta\dot{Q}F''' +\frac{\theta\dot{Q}F''}{2Q}-\frac{\theta\dot{Q}F'}{2Q^2}\Bigg](1+w)\mathcal{W}^k  +\left[ \frac{\theta F'}{2Q} -\theta F''\right](1+w)\dot{\mathcal{W}}^k\,, \nonumber\\&&
		\ddot{\mathcal W}^k=\frac{\dddot{Q}}{\dot{Q}}\mathcal W^k-\frac{2w \ddot{Q}}{1+w}\delta^k_m-\frac{w \dot{Q}}{1+w}\dot{\delta}_m^k\,.\label{700x}
	\end{eqnarray}
 }
 As extensively used in numerous modified gravity studies in \cite{sahlu2020scalar, sami2021covariant, Ntahompagaze, abebe2013large}, the quasi-static approximation is a robust approach in $f(Q)$ gravity 
 \cite{sahlu2025structure}. We adopted this approximation, wherein the first and second-order time derivatives of non-metric density fluctuations are presumed to be nearly zero ($\dot{\mathcal{W}} = \ddot{\mathcal{W}} \approx 0$). Under this approximation, Eqs. \eqref{70011} - \eqref{700x} simplify to a closed system of equations for a matter-dominated Universe as
\begin{eqnarray}
		&&\ddot{\delta}_m^k= -\left( 2H+\frac{\dot{H}}{H}F'+12H\dot{H}F'' \right)\dot{\delta}_m^k+\frac{\rho_m}{2}\delta^k_m \,. \label{70011x}
\end{eqnarray}
For the case of $F' = F'' = 0$, the above equation \eqref{70011x}  reduces to $\Lambda$CDM limit. 
By admitting the redshift-space transformation technique so that any first-order and second-order time derivative functions ${y(t)}$ become $\dot{y} = -(1+z)H\frac{{\rm d} y}{{\rm d}z}$, and the second time derivative becomes $$\ddot{y} = (1+z)H^2\frac{{\rm d}y}{{\rm d}z} +(1+z)^2H^2\frac{{\rm d}^2y}{{\rm d}z^2}+  (1+z)^2H\frac{{\rm d}H}{{\rm d}z} \frac{{\rm d}y}{{\rm d}z}\;.$$ Then, the redshift-space transformation of Eq. \eqref{70011x} yields
\begin{eqnarray}\label{densitycontrast123}
   && \frac{{\rm d}^2\delta_m}{{\rm d}z^2} =    \left(\frac{1}{1+z}- \frac{{\rm d}E}{E{\rm d}z}\left(1 +\bar{f}_{1,2,3} \right)\right)\frac{{\rm d} \delta_m^k}{{\rm d}z}\nonumber\\&& +\frac{\Omega_{m}}{2E^2}(1+z)^{1-3\zeta}\delta^k_m \;,
\end{eqnarray}
where $\bar{f}_{1,2,3}  \text{ is represent } n(1-\Omega_m)$, $\frac{(1-\Omega_m)p\,{\rm e}^{-p}}{1-(1+p)\,{\rm e}^{-p}}$ and $\frac{3(1-\Omega_m)}{2-\ln(\Gamma)}$ corresponding results for $f_{1\rm{CDM}}$, $f_{2\rm{CDM}}$, and $f_{3\rm{CDM}}$ models respectively. 
\\
\\
The growth factor represented by $D(z)$ is the ratio of the amplitude of density contrast at an arbitrary redshift $z_{i}$ and is expressed as 
\begin{eqnarray}
     D(z) = \frac{\delta(z)}{\delta(z_{i})}\;,
 \end{eqnarray}
The growth factor quantifies the amplitude of density perturbations at any given time relative to their initial values. It is often normalized to be $\delta(z_{i}) = 1$ at present ($z =0$). Mathematically, the growth factor's evolution is governed by differential equations that include the cosmological expansion rate\cite{percival2005cosmological}. This quantity is essential for understanding how small initial overdensities in the matter distribution grow due to gravitational attraction, leading to the formation of large-scale structures. The growth factor's dependence on the Universe's composition, including matter and dark energy, highlights how these components influence structure formation. In a dark energy-dominated Universe, the structure growth slows as the accelerated expansion counteracts gravitational collapse. The growth factor is vital for modeling galaxy formation and large-scale structures, and for comparing theoretical predictions with observations from the CMB, galaxy surveys, and other large-scale structure surveys. Additionally, the related growth rate $f(z)$, which measures the rate at which structures grow, is used in various observational probes, including redshift-space distortions ${f}\sigma_8$. The growth rate $f(z)$, as obtained from the density contrast $\mathcal{\delta}_m$, yields \cite{springel2006large}
\begin{equation}
    f 
    \equiv       
    \frac{{\rm d}\ln{{{\delta}}_m}}{{\rm d}\ln{a}} = -(1+z)\frac{\delta'_m(z)}{\delta_m(z)}  
    \;.\label{growth1}
\end{equation}
Therefore, the growth factor is fundamental in understanding the dynamical evolution of the Universe's structures. By substituting the definition of \eqref{growth1} into the second-order evolution Eq. \eqref{densitycontrast123}, the evolution of the growth rate is governed by the following expression
%
%
\begin{eqnarray}\label{growthrate}
     (1+z)f' =f^2 +\Big[ 2 -(1+z)\frac{dE}{Edz}\bar{f}_{1,2,3}  \Big]f  -\frac{\Omega_{m}}{2E^2}(1+z)^{1-3\zeta}\;. \nonumber\\&&
\end{eqnarray}

%
As presented in \cite{avila2022inferring,adil2024s}, a good approximation of the growth rate $f(z)$ is
\begin{eqnarray}
    f(z) = \tilde{\Omega}^\gamma_m(z)\;,
\end{eqnarray}
where $\tilde{\Omega}(z) \equiv \frac{\Omega_m(z)}{H^2(z)/H^2_0}$, and $\gamma$ being the growth index. As presented in \cite{linder2007parameterized}, the $w$CDM growth index seems to satisfy, $\gamma = 3(w-1)/(6w-5)$. For the case of $w = -1$, the $\Lambda$CDM model results in the value of $\gamma = 6/11$, but this varies for different alternative gravity models. A combination of the linear growth rate $f(z)$ with the root mean square normalization of the matter power spectrum $\sigma_8$ within the radius sphere $8h^{-1}\,\mathrm{Mpc}$, yields the redshift-space distortion $f\sigma_8$ and the detailed clarification is presented in the work \cite{avila2022inferring,adil2024s}.

\section{Data and Methodology}\label{datameth}
The recent cosmological measurements, namely: \textit{CC}, \textit{BAO}, \textit{PantheonP + SH0ES},  $f\sigma_8$, and \textit{f}  together with a joint analysis of {\textit{CC} + \textit{BAO} }, \textit{f} + $f\sigma_8$, {\textit{PantheonP + SH0ES} + \textit{f}} and { \textit{PantheonP + SH0ES} + $f\sigma_8$} + {$f\sigma_8$ + \textit{f}} have been considered for an in-depth observational and statistical analysis.

\begin{enumerate} 
\item \textit{CC}: The Hubble parameter $H(z)$ measurements, which have 31 data points derived from the relative ages of massive, early-time, passively evolving galaxies, known as cosmic chronometers. Hereafter, the dataset is labeled as CC. We calculate the minimum $\chi^2$ with a cosmic chronometer covariance in combination with a statistical and systematic errors as  clarified in \cite{moresco2020setting,qi2023model}
\begin{equation}
        \chi^2_{\text{CC}} = \left(H_{\text{theo}}(z)-H_{\text{ob}}(z_i)\right)^{T}C^{-1}\left(H_{\text{theo}}(z)-H_{\text{ob}}(z_i)\right)\;,
    \end{equation}
    where $H_{\text{theo}}(z)$ stands for each theoretical model and $H_{\text{ob}}(z_i)$ refers to the Hubble parameter data.
    
\item  \textit{BAO:}  We use \textit{BAO} data from distance and the correlation measurements of DESI \cite{adame2024desi}.  The measurements include data for the isotropic \textit{BAO} measurements of $D_V(z)/r_d$, where $D_V(z)$ and $r_d$ are the spherically averaged volume distance and sound horizon at baryon drag respectively, anisotropic \textit{BAO} measurements of $D_M(z)/r_d$ and $D_H(z)/r_d$, where $D_M(z)$ and $D_H(z)$ are the comoving angular diameter distance and the Hubble distance, respectively, and correlations between them. Hereafter, we refer to this dataset as $BAO$. The general form of distance modulus \(\mu(z)\) has been taken into consideration as 
 \begin{equation}\label{distancemodules1}
 	\mu(z) =  25+5\log_{10}{D_L(z)}\;,
 \end{equation}
 since $D_L$ is the luminosity distance, and it yields 
 \begin{equation}
 \label{D_L}
     D_{L}(z) = 3000\bar{h}^{-1}(1+z)\int^{z}_{0}\frac{{\rm d}{z'}}{E(z')}\;,
 \end{equation}
  where $\bar{h} = (H_0\,\mathrm{km/s/Mpc)}/100$. The volume-averaged angular diameter distance reflects \textit{BAO} measurements averaged over spherical distances 
\begin{equation}
 D_{V}(z)=\left[(1+z)^2 D^2_{A}(z)^{2}\frac{300\bar{h}^{-1}z}{E(z)}\right]^{\frac{1}{3}}~,   
\end{equation}
and the angular distance is yielded as   
\begin{equation}
D_{A}(z)=\frac{\bar{h}^{-1}}{(1+z)}\int_{0}^{z}\frac{{\rm d}z'}{E(z')}\;.
\end{equation}
The sound horizon at the drag epoch is defined as
\begin{equation}
r_d=\int_{z_d}^{\infty}\frac{c_s(z)}{(H_0\,\mathrm{km/s/Mpc})E(z)} dz~,   
\end{equation}
where $z_d$ is the redshift at drag epoch and $c_s(z)$ is the photon-baryon fluid's sound speed.
The  minimum $BAO$ $\chi^2$ for the \textit{BAO} dataset is given by:
    \begin{eqnarray}
        \chi^2_{\it{BAO}} = (\mathbf{D_{\text{theo}}}-\mathbf{D_{\text{obs}}} )^T \mathbf{C}^{-1} (\mathbf{D_{\text{theo}}}-\mathbf{D_{\text{obs}}} )\;,
    \end{eqnarray}
    where \( \mathbf{D_{\text{obs}}} \) is the vector of observed measurements and \( \mathbf{D_{\text{theo}}} \) is the vector of theoretical model predictions.
\item \textit{{Pantheon Plus + SH0ES} sample:}  
    We use the SNe Ia distance moduli measurements from the \textit{Pantheon Plus} sample \cite{brout2022Pantheon+}, which consists of 1701 light curves of 1550 distinct SNe Ia ranging in the redshift interval $z \in [0.001, 2.26]$. The best-fit values of the cosmological parameters are constrained by minimizing a $\chi^2$ likelihood as presented in \cite{brout2022Pantheon+}
    \begin{eqnarray}
        \chi^2 = \Delta \vec{D}
        ^T_i(z) C^{-1}\Delta \vec{D}
        _{i}(z)\;,
    \end{eqnarray}
 where $C$ is the statistical and systematic covariance matrices used for constraining cosmological models.  $\vec{D}$ is the vector of 1701 SNIa distance modulus residuals that are computed as $\Delta \vec{D}_i = \mu_i-\mu_{model}(z_i)$, i.e. each SNIa distance $\mu_i$ is derived from the observed (and corrected) apparent magnitude and a calibrated absolute magnitude ($M_{abs}$). The residual $\Delta \vec{D}_i$ then represents the difference between these observed values and the distance modulus predicted by the cosmological model.

 The two parameters $M_{abs}$ and $H_0\,\mathrm{(km/s/Mpc)}$ are degenerate when evaluating SNIa alone, as stated in \cite{brout2022Pantheon+}. This degeneracy stems from the common assumption that SNIa are standard candles with a fixed absolute magnitude, implying that the observed distance modulus is determined solely by the apparent magnitude. In practice, however, there are slight intrinsic variations in the luminosity of SNIa, meaning they are only standard, not perfect, candles. To break this degeneracy, SNIa observations are calibrated using Cepheid variable stars, which serve as true standard candles and provide accurate distances to the host galaxies. {In the \textit{Pantheon Plus} analysis, recent \textit{SH0ES} Cepheid host distances are incorporated into the likelihood. Thus, in the following  we shall refer to this dataset as {\textit{PantheonP + SH0ES}}}. Accordingly, the modified SNIa distance residual $\Delta D'$ is redefined as follows
\[
\Delta D' = 
\begin{cases} 
\mu_i - \mu_i^{\text{Cepheid}} & \text{if } i \text{ is the Cepheid host} \\ 
\mu_i - \mu_{\text{model}}(z_i) & \text{otherwise,} 
\end{cases}
\]
where $\mu_i^{\text{Cepheid}}$ is the Cepheid calibrated host galaxies distance provided by \textit{SH0ES}. The residual $\mu_i - \mu_i^{\text{Cepheid}}$ is sensitive to both $M_{abs}$ and $H_0\,\mathrm{(km/s/Mpc)}$. To further reduce any discrepancies, we allow $M_{abs}$ to be a free parameter. Admittedly, this adds an extra free parameter, but it enables a more robust constraint on $H_{0}$ by avoiding potential biases associated with fixing $M_{abs}$ at a fixed value (e.g., -19.2) that may not be representative for all SNIa. The  $\chi^2$ used for the \textit{PantheonP + SH0ES} dataset is therefore defined as: 
\begin{eqnarray}
\chi^2_{\text{\textit{PantheonP + SH0ES}}} &=& \Delta D'^T \left( C^{\text{SNIa}}_{\text{syst+stat}}+C^{\text{Cepheid}}_{\text{syst+stat}}\right)^{-1}\Delta D'\nonumber\\
&&\;.
\end{eqnarray}

\item  We also incorporate redshift-space distortion data, labeled $f\sigma_8$, and the latest growth rate data, labeled \textit{f} from the VIMOS Public Extragalactic Redshift Survey (\textit{VIPERS}) and SDSS collaborations. Specifically, we utilize:
\begin{enumerate}
\item A total of 66 data points of the measurements of redshift-space distortion for $f\sigma_8$ have been collected and summarized in the works of  \cite{skara2020tension,kazantzidis2018evolution}, covering the redshift interval $0.001 \leq z \leq 1.944$.
    \item Additionally, there are 14 data points for \textit{f} within the redshift range $0.001 \leq z \leq 1.4$ which are presented in \cite{avila2022inferring,Perenon}. The resulting  $\chi^2$ for redshift-space distortion and growth rate is expressed by a standard Gaussian distribution as:
\end{enumerate}
\begin{eqnarray}
\chi^2_{f\sigma_8} = \frac{\left(f\sigma_{8\,\text{theo}}(z)-f\sigma_{8\,\text{ob}}(z_i)\right)^{2}}{\sigma^2_{f\sigma_8}}
\end{eqnarray}  
and
\begin{eqnarray}
\chi^2_{f} = \frac{\left(f_{\text{theo}}(z)-f_{\text{ob}}(z_i)\right)^{2}}{\sigma^2_f}\;,
\end{eqnarray}    
respectively.
\item We also used a combined analysis of the above measurements and their  $\chi^2$ for these joint datasets yields
\begin{eqnarray}
\chi^2_{\text{tot}} =  \sum_i \chi^2_{\text{i}}\,,
\end{eqnarray}
where $i$ represent {\it CC}, {\it{BAO}}, {\text{\textit{PantheonP + SH0ES}}}, ${f\sigma_8}$, and \textit{f}. By considering the inconsistency of the \textit{PantheonP +SH0ES} \& $BAO$ datasets as presented in the  \cite{afroz2025hint} disclosed through a violation of the distance duality relation, we have only consider the joint analysis $CC$+$BAO$, \textit{ PantheonP+SH0ES + $f\sigma_8$} and  \textit{PantheonP+SH0ES +f}.  
\item In this work, we use the newly released MCMC simulation Python package, called \texttt{Kosmulator}, specifically designed for constraining modified gravity models on cosmological datasets\footnote{The \texttt{Kosmulator} installation documentation can be found at: \url{https://github.com/renierht/Kosmulator}}. \texttt{Kosmulator} utilize the Python packages \texttt{EMCEE} \cite{foreman2013emcee} and \texttt{GetDist} \cite{lewis2019getdist} to constrain and plot the model parameters, namely: \(\Omega_m\), $H_0\,\mathrm{(km/s/Mpc)}$, \(r_{d}\), \(M_{abs}\), \(n\), \(p\), \(\Gamma\), \(\sigma_8\), \(\zeta\) {and} \(\gamma\) using these cosmological datasets. 
\end{enumerate}

\section{Results and Discussion}\label{resultdiscussion}
\subsection{Constraining parameters}
In this section, a detailed, comprehensive analysis of background evolution and structure growth is showcased for $f(Q)$ gravity models with and without viscous effects. As aforementioned, in the following we shall utilize \texttt{Kosmulator}, an updated MCMC simulation originally developed in \cite{hough2020viability}, to estimate the best-fit Gaussian distribution values of the cosmological parameters:  \{$\Omega_m$, $H_0\,\mathrm{(km/s/Mpc)}$, $r_d$, $M_{abs}$, $\zeta$, $\gamma$, $\sigma_8$\}, and the exponents \{$n$, $p$, $\Gamma$\} for the three $f(Q)$ gravity models. 
\begin{table*}[h!]
\renewcommand{\arraystretch}{1.5}  
\centering
\begin{tabular}{l @{\hskip 6pt} r @{\hskip 6pt} r @{\hskip 6pt} r @{\hskip 6pt} r @{\hskip 6pt} r @{\hskip 6pt} r @{\hskip 6pt} r @{\hskip 6pt} r @{\hskip 6pt} r}
\hline
\textbf{Data} & \textit{CC}&\textit{BAO} & {\textit{PantheonP }} & \textit{CC} + \textit{BAO} & {$f\sigma_8$} & \textit{f} & \textit{f} + $f\sigma_8$ & {\textit{PantheonP}} & {\textit{PantheonP }} \\ 
 &  &  & \textit{+ SH0ES} &  &  &  &  & {\textit{ + SH0ES} + \textit{f} } & {\textit{+ SH0ES} + $f\sigma_8$} \\
\hline
&$f_1$CDM&&&&&&\\
$H_0$ &$67.320^{+2.953}_{-2.759}$   & $69.179^{+7.200}_{-6.426}$ & $73.534^{+1.012}_{-1.003}$ &  $68.184^{+1.812}_{-1.830}$& -- & -- & -- &  $73.548^{+1.031}_{-1.000}$  &$73.534^{+1.026}_{-1.005}$  \\ 
$\Omega_m$ &$0.297^{+0.056}_{-0.055}$&$0.288^{+0.016}_{-0.017}$ & $0.284^{+0.036}_{-0.041}$&   $0.289^{+0.016}_{-0.016}$  & $0.261^{+0.067}_{-0.072}$  & $0.266^{+0.051}_{-0.061}$ & $0.263^{+0.054}_{-0.064}$& $0.292^{+0.024}_{-0.032}$ & $0.280^{+0.035}_{-0.038}$ \\ 
$n$ &$0.246^{+0.170}_{-0.166}$ & $0.168^{+0.164}_{-0.117}$   & $0.245^{+0.169}_{-0.164}$& $0.164^{+0.161}_{-0.114}$  &$0.263^{+0.161}_{-0.174}$ & $0.183^{+0.175}_{-0.129}$&$0.202^{+0.161}_{-0.137}$  & $0.200^{+0.144}_{-0.121}$  & $0.264^{+0.154}_{-0.167}$\\ 
$r_d$ &-- & $145.163^{+14.845}_{-13.659}$  & -- & $147.178^{+3.534}_{-3.352}$ &  --& -- & -- &  --& --\\ 
$M_{abs}$ &-- & --&  $-19.243^{+0.029}_{-0.029}$&  -- & -- & --&-- & $-19.244^{+0.030}_{-0.029}$ & $-19.243^{+0.029}_{-0.029}$\\ 
{$\sigma_8$} & --&-- & --& & $0.810^{+0.045}_{-0.038}$ &--   & $0.807^{+0.026}_{-0.023}$ &--& $0.803^{+0.044}_{-0.036}$\\ 
$s_8$ & -- & -- & --&-- & $0.752^{+0.099}_{-0.099}$ & -- & $0.754^{+0.081}_{-0.081}$ & --&$0.776^{+0.064}_{-0.063}$ \\ 
$\gamma$ & -- & -- & --&-- &$0.552^{+0.098}_{-0.100}$& $0.575^{+0.087}_{-0.106}$  & $0.563^{+0.094}_{-0.102}$ & $0.619^{+0.054}_{-0.076}$  & $0.580^{+0.063}_{-0.063}$    \\ 
\hline
&$f_2$CDM&&&&&&\\
$H_0$ & $70.211^{+3.061}_{-3.103}$ &  $69.540^{+7.003}_{-6.540}$   &$73.752^{+1.010}_{-1.002}$&$70.931^{+1.918}_{-1.876}$ & --&--&-- & $73.813^{+1.013}_{-1.008}$& $73.762^{+1.023}_{-1.002}$ \\ 
$\Omega_m$ &$0.280^{+0.056}_{-0.049}$  & $0.263^{+0.014}_{-0.012}$ & $0.321^{+0.034}_{-0.024}$& $0.236^{+0.013}_{-0.012}$  & $0.260^{+0.072}_{-0.068}$   & $0.219^{+0.058}_{-0.051}$  & $0.248^{+0.062}_{-0.051}$ &  $0.307^{+0.018}_{-0.017}$ &  $0.309^{+0.021}_{-0.018}$\\ 
$p$ & $0.842^{+0.110}_{-0.153}$ & $0.918^{+0.054}_{-0.063}$  &$0.918^{+0.059}_{-0.098}$&$0.926^{+0.049}_{-0.061}$ & $0.827^{+0.127}_{-0.405}$ &  $0.694^{+0.172}_{-0.154}$  & $0.869^{+0.090}_{-0.120}$ & $0.948^{+0.037}_{-0.057}$  & $0.949^{+0.037}_{-0.065}$ \\ 
$r_d$ & -- & $150.934^{+15.705}_{-13.859}$ & --&  $147.501^{+3.553}_{-3.385}$  & -- & -- & -- &  -- & --\\ 
$M_{abs}$ & -- & -- & $-19.245^{+0.029}_{-0.030}$&-- & -- &-- & -- &  $-19.245^{+0.029}_{-0.030}$ &$-19.245^{+0.029}_{-0.030}$  \\ 
$\sigma_8$ &-- & -- & --&-- & $0.773^{+0.046}_{-0.115}$ & -- & $0.798^{+0.025}_{-0.025}$ & --& $0.790^{+0.025}_{-0.027}$\\ 
{$s_8$}& -- & -- & --&-- & $0.7203^{+0.02}_{-0.01}$ & -- & $0.7245^{+0.113}_{-0.097}$ & -- & $0.802^{+0.037}_{-0.037}$\\
$\gamma$ & --& -- & --&-- & $0.487^{+0.125}_{-0.067}$     
 & $0.536^{+0.105}_{-0.088}$  & $0.517^{+0.112}_{-0.084}$ &   $0.651^{+0.034}_{-0.049}$   &$0.632^{+0.045}_{-0.059}$  \\ 
\hline
&$f_3$CDM&&&&&&\\
$H_0$ & $70.708^{+2.633}_{-2.397}$ & $69.403^{+7.032}_{-6.551}$ & $73.768^{+1.008}_{-1.005}$ &  $71.179^{+1.784}_{-1.780}$ & --&--&-- & $73.920^{+1.020}_{-1.000}$ & $73.809^{+1.015}_{-1.002}$ \\ 
$\Omega_m$ &$0.352^{+0.033}_{-0.045}$ &$0.310^{+0.015}_{-0.014}$  &$0.388^{+0.009}_{-0.014}$ & $0.313^{+0.015}_{-0.014}$  & $0.278^{+0.070}_{-0.065}$  &  $0.245^{+0.051}_{-0.048}$   &$0.258^{+0.058}_{-0.043}$ & $0.359^{+0.016}_{-0.016}$ & $0.384^{+0.011}_{-0.015}$ \\ 
$\Gamma$ & $3.457^{+1.269}_{-1.014}$ &  $2.704^{+0.826}_{-0.506}$&  $2.459^{+0.531}_{-0.330}$&  $2.661^{+0.768}_{-0.479}$   & $3.887^{+1.500}_{-1.284}$   & $3.751^{+1.825}_{-1.243}$ &  $3.009^{+1.156}_{-0.728}$& $2.208^{+0.306}_{-0.155}$& $2.410^{+0.498}_{-0.297}$\\ 
$r_d$ & -- & $153.498^{+14.856}_{-14.461}$ & -- & $147.304^{+3.506}_{-3.343}$& -- &--& -- & -- & -- \\ 
$M_{abs}$ &-- & -- & $-19.246^{+0.029}_{-0.029}$&-- &--& --&-- &  $-19.247^{+0.030}_{-0.029}$&  $-19.246^{+0.029}_{-0.029}$  \\ 

$\sigma_8$ & -- & -- & -&-- &  $0.720^{+0.037}_{-0.042}$ & -- & $0.772^{+0.024}_{-0.023}$  &--&   $0.709^{+0.016}_{-0.016}$   \\ 
{$s_8$} & -- & --& -&-- & $0.693^{+0.123}_{-0.121}$ & --  & $0.715^{+0.102}_{-0.083}$& --& $0.802^{+0.020}_{-0.020}$ \\ 
$\gamma$ & -- &-- & --&-- &  $0.536^{+0.105}_{-0.091}$ &$0.518^{+0.117}_{-0.077}$  & $0.463^{+0.088}_{-0.046}$ & $0.685^{+0.011}_{-0.023}$    & $0.648^{+0.036}_{-0.053}$     \\
\hline
&$\Lambda$CDM&&&&&&\\
$H_0$ & $68.335^{+2.833}_{-2.542}$ &  $69.116^{+7.175}_{-6.318}$  &$73.588^{+1.002}_{-1.001}$&   $69.181^{+1.714}_{-1.697}$  & --&--&-- &$73.684^{+1.012}_{-1.002}$ &$73.592^{+1.032}_{-0.991}$\\ 
$\Omega_m$ &$0.316^{+0.049}_{-0.050}$ &$0.294^{+0.015}_{-0.014}$&$0.331^{+0.018}_{-0.018}$&   $0.296^{+0.015}_{-0.014}$ &   $0.270^{+0.068}_{-0.071}$ &  $0.257^{+0.047}_{-0.059}$ & $0.250^{+0.056}_{-0.058}$ &  $0.321^{+0.015}_{-0.015}$ &  $0.328^{+0.017}_{-0.017}$ \\  
$r_d$ & -- &  $147.488^{+14.894}_{-13.814}$ & --&$147.197^{+3.496}_{-3.386}$& -- & -- & -- &   -- & -- \\ 
$M_{abs}$ & -- & -- & $-19.244^{+0.029}_{-0.029}$&--& -- & --  & -- &  $-19.244^{+0.029}_{-0.029}$& $-19.245^{+0.030}_{-0.029}$ \\ 
$\sigma_8$ & -- & -- & --&-- & $0.776^{+0.033}_{-0.028}$ & -- & $0.792^{+0.023}_{-0.021}$& --&  $0.760^{+0.020}_{-0.020}$\\ 
{$s_8$} & -- & -- & --&-- &$0.736^{+0.031}_{-0.035}$ & -- & $0.737^{+0.025}_{-0.024}$ & --& $0.795^{+0.029}_{-0.029}$\\ 
$\gamma$ & -- &-- & -- &--  & $0.540^{+0.102}_{-0.095}$  &  $0.577^{+0.086}_{-0.107}$     
 & $0.544^{+0.099}_{-0.094}$ & $0.667^{+0.024}_{-0.039}$   &  $0.621^{+0.050}_{-0.059}$\\ 
\hline
\end{tabular}
\caption{{{The calculated best-fit parameters through MCMC simulations in the $\zeta = 0$, for $\Lambda$CDM, $f_{1\rm CDM}$, $f_{2\rm CDM}$ and $f_{3\rm CDM}$ models. Note that $H_{0}$ is given in $\mathrm{km/s/Mpc}$.}}}
\label{tab:best_fit_valueswithout}
\end{table*}

\begin{figure*}[h!]
\centering
 		\includegraphics[scale=0.3]{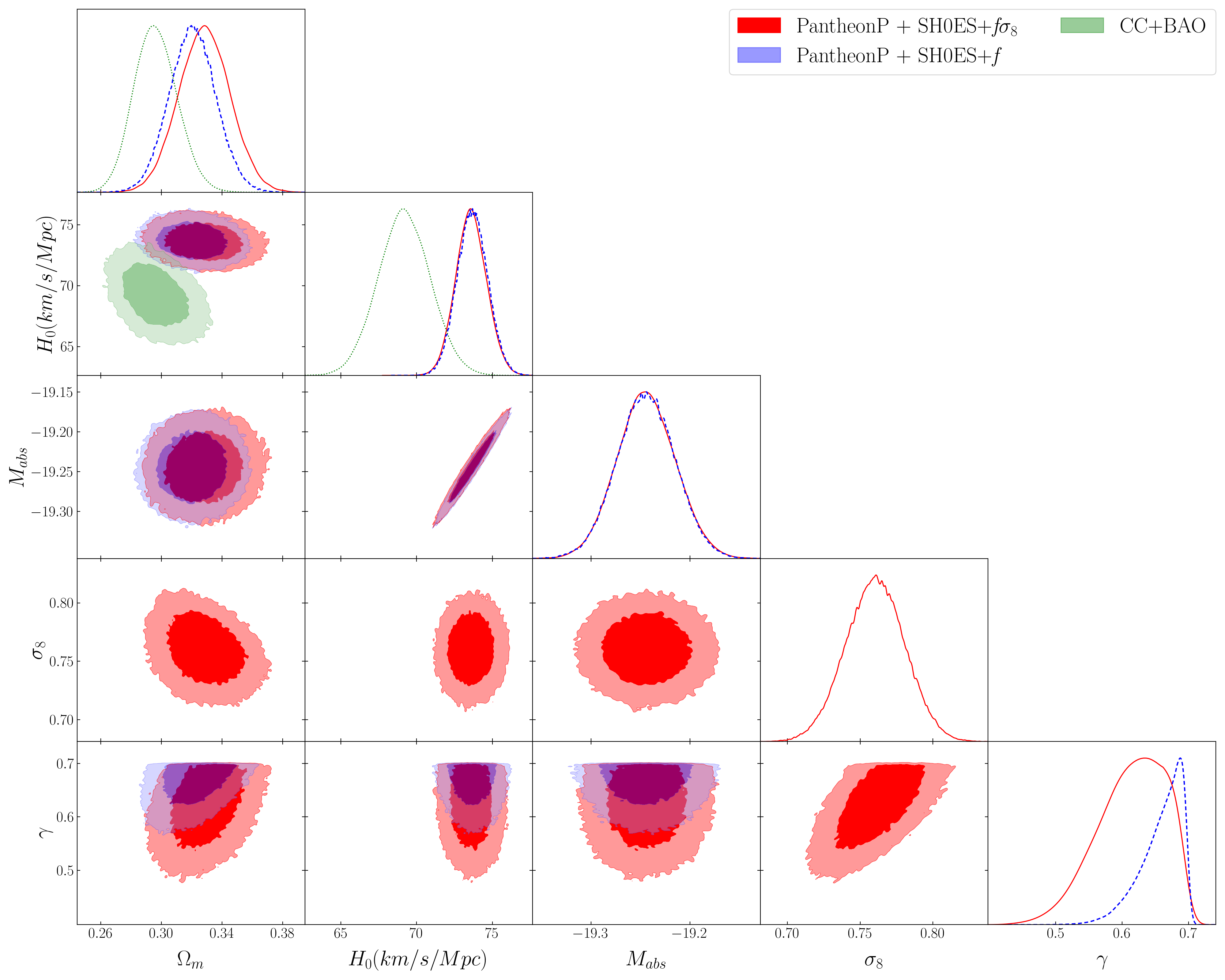}
   \caption{  Best-fit parameter values at $1\sigma$ and $2\sigma$ confidence levels for $\Lambda$CDM in the case of non-viscous term ($\zeta = 0$) using the  {\textit{PantheonP + SH0ES} + $f\sigma_8$}, { \textit{PantheonP + SH0ES} +$f$}, and {\textit{CC}+ \textit{BAO}} datasets. }
    \label{fig:enterlcdm1}
\end{figure*}

\begin{figure*}[h!]
\centering
 		\includegraphics[scale=0.3]{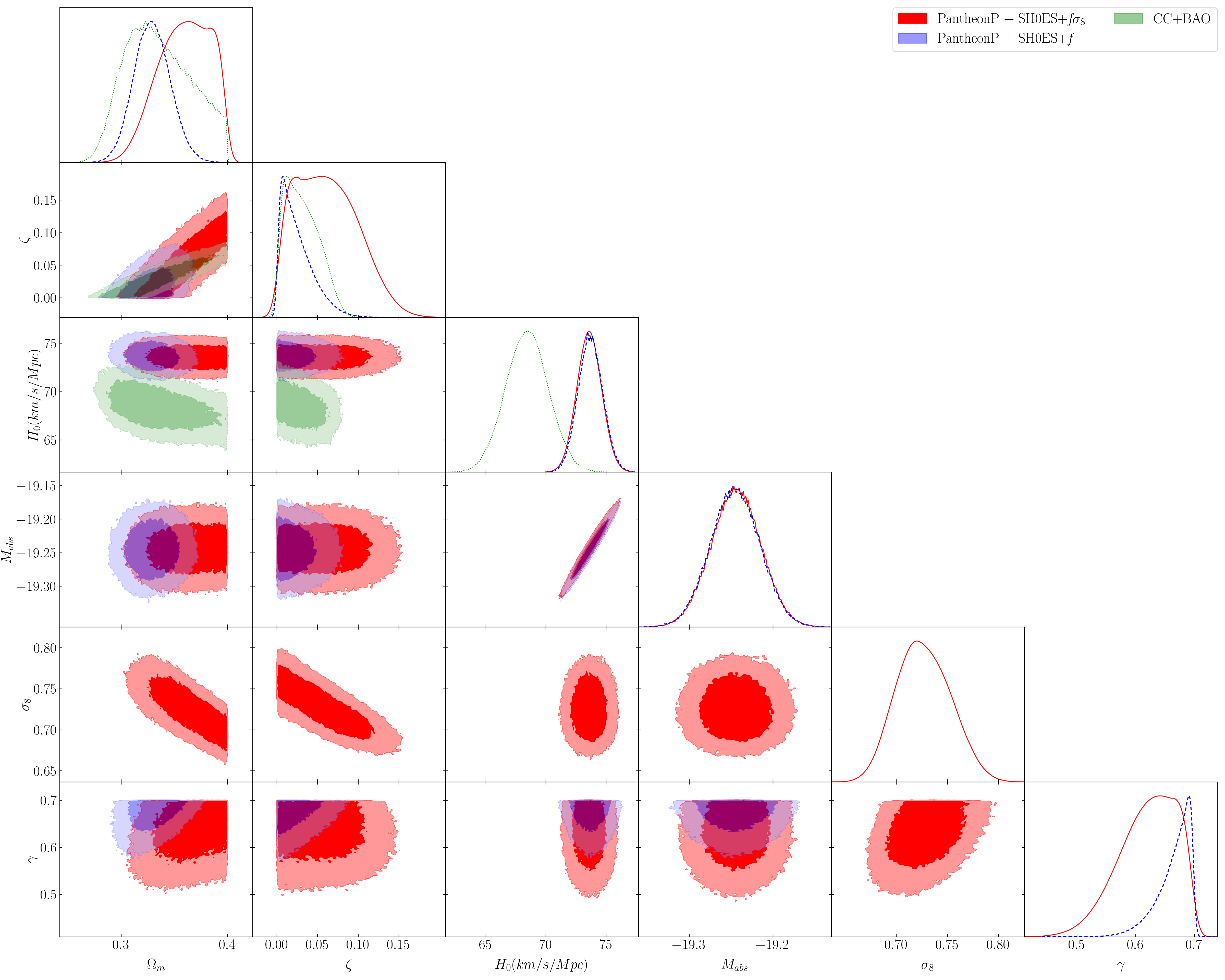}
   \caption{ Best-fit parameter values at $1\sigma$ and $2\sigma$ confidence levels for $\Lambda$CDM in the case of a viscous term ($\zeta > 0$) using the {\textit{PantheonP + SH0ES} + $f\sigma_8$}, { \textit{PantheonP + SH0ES} +$f$}, and {\textit{CC}+ \textit{BAO}} datasets. }
    \label{fig:enterlcdm2}
\end{figure*}

\begin{figure*}[h!]
            \centering
 		\includegraphics[scale=0.3]{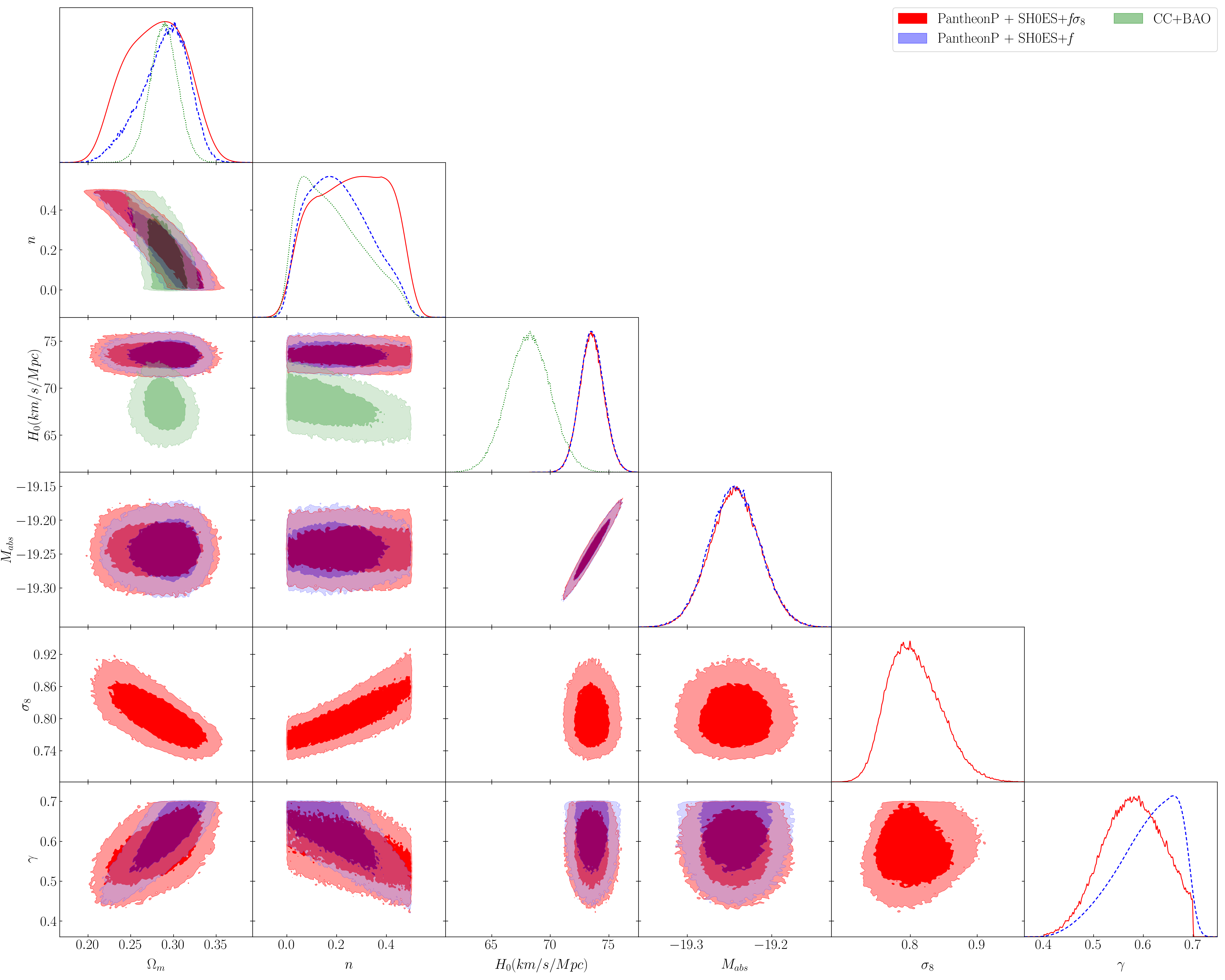}
   \caption{Best-fit parameter values for $f_{1\rm CDM}$ in the case of ($\zeta = 0$) using the datasets {\textit{PantheonP + SH0ES} + $f\sigma_8$}, { \textit{PantheonP + SH0ES} +$f$}, and {\textit{CC}+ \textit{BAO}}, evaluated at $1\sigma$ and $2\sigma$ confidence levels.}
    \label{fig:enterpower}   
\end{figure*}

\begin{figure*}[h!]
 		\centering
 		\includegraphics[scale=0.25]{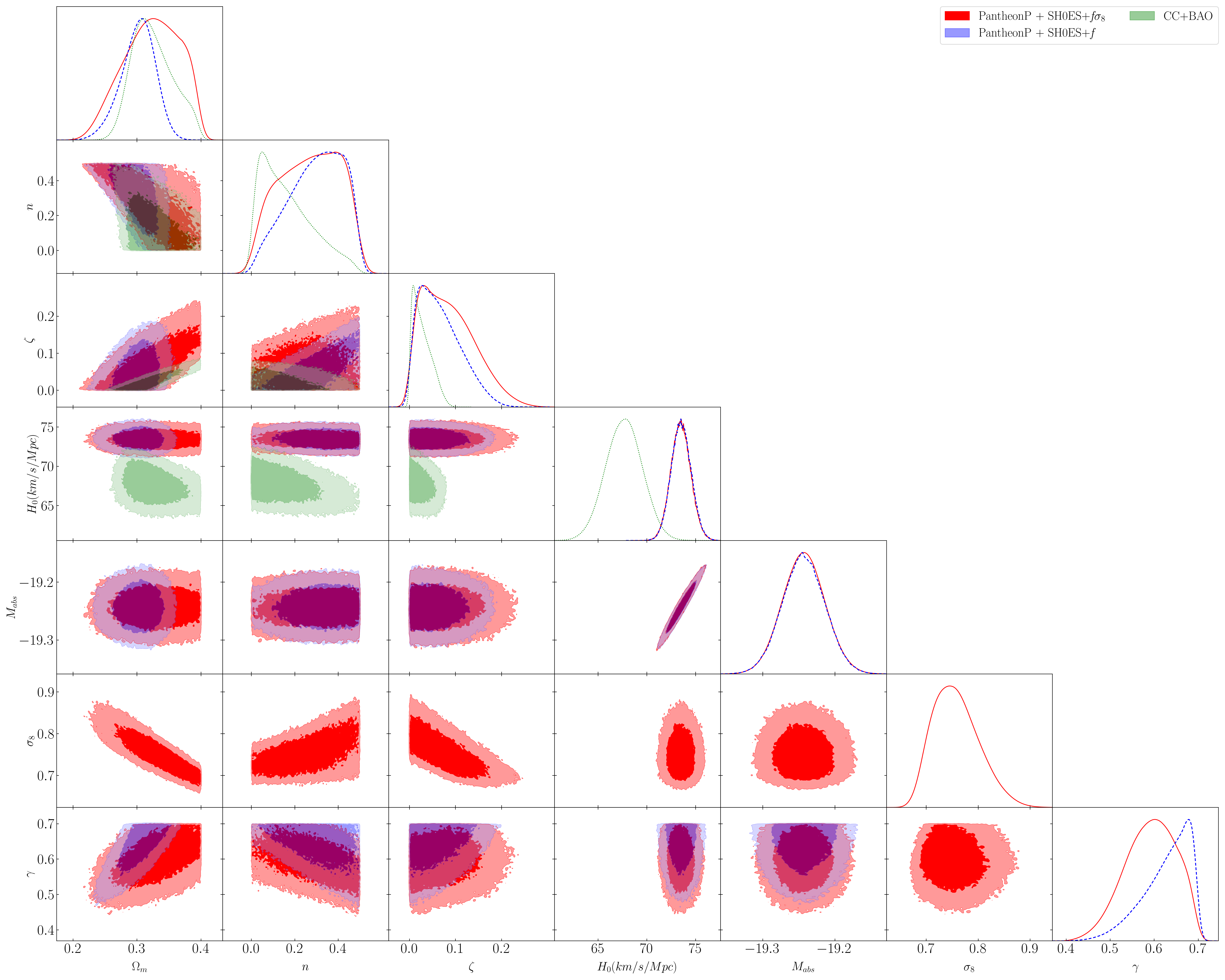}
   \caption{Best-fit parameter values for $f_{1\rm CDM}$ in the case of ($\zeta \neq 0$), using the datasets {\textit{PantheonP + SH0ES} + $f\sigma_8$}, { \textit{PantheonP + SH0ES} +$f$}, and {\textit{CC}+ \textit{BAO}}, evaluated at $1\sigma$ and $2\sigma$ confidence levels.}
    \label{fig:enterpower1}   
\end{figure*}

\begin{table*}[h!]
\renewcommand{\arraystretch}{1.5}  
\centering
\begin{tabular}{l @{\hskip 6pt} r @{\hskip 6pt} r @{\hskip 6pt} r @{\hskip 6pt} r @{\hskip 6pt} r @{\hskip 6pt} r @{\hskip 6pt} r @{\hskip 6pt} r @{\hskip 6pt} r}
\hline
\textbf{Data} & \textit{CC}&\textit{BAO} & {\textit{PantheonP }} & \textit{CC} + \textit{BAO} & {$f\sigma_8$} & \textit{f} & \textit{f} + $f\sigma_8$ & {\textit{PantheonP}} & {\textit{PantheonP }} \\ 
 &  &  & \textit{+ SH0ES} &  &  &  &  & {\textit{ + SH0ES} + \textit{f} } & {\textit{+ SH0ES} + $f\sigma_8$} \\
\hline
&$f_1$CDM&&&&&&\\
$H_0$ & $67.907^{+2.550}_{-2.427}$ & $68.677^{+7.498}_{-5.756}$ &  $73.467^{+1.020}_{-0.996}$& $67.662^{+1.854}_{-1.875}$ & --&--&--& $73.508^{+1.029}_{-0.994}$  &$73.472^{+1.016}_{-0.997}$\\ 
$\Omega_m$ & $0.335^{+0.045}_{-0.061}$  & $0.321^{+0.041}_{-0.029}$ & $0.332^{+0.043}_{-0.050}$&  $0.321^{+0.039}_{-0.028}$& $0.294^{+0.068}_{-0.077}$ & $0.289^{+0.013}_{-0.015}$ & $0.243^{+0.052}_{-0.054}$ &   $0.303^{+0.024}_{-0.028}$ &  $0.326^{+0.045}_{-0.048}$ \\ 
$n$ &$0.261^{+0.167}_{-0.173}$  & $0.135^{+0.160}_{-0.100}$ & $0.256^{+0.160}_{-0.163}$& | $0.139^{+0.155}_{-0.100}$  & $0.241^{+0.164}_{-0.190}$ & $0.236^{+0.172}_{-0.160}$  &  $0.284^{+0.147}_{-0.176}$&   $0.311^{+0.122}_{-0.149}$ &  $0.276^{+0.147}_{-0.166}$\\ 
$r_d$ &-- & $145.178^{+13.205}_{-14.586}$  & -- &   $147.171^{+3.542}_{-3.339}$  & -- & -- & -- & --&-- \\ 
$M_{abs}$ &-- & -- & $-19.244^{+0.030}_{-0.029}$ &  --  & -- &-- & -- &  $-19.244^{+0.030}_{-0.029}$ &$-19.243^{+0.029}_{-0.029}$   \\ 
$\zeta$ & $0.054^{+0.060}_{-0.037}$ & $0.024^{+0.027}_{-0.017}$  &$0.080^{+0.066}_{-0.054}$ & $0.024^{+0.026}_{-0.017}$   & $0.133^{+0.105}_{-0.091}$  &  $0.151^{+0.095}_{-0.093}$ & $0.085^{+0.083}_{-0.057}$ & $0.059^{+0.055}_{-0.040}$&  $0.077^{+0.067}_{-0.054}$ \\ 
$\sigma_8$ & --& --  & -- & -- & $0.747^{+0.055}_{-0.048}$ & -- & $0.790^{+0.027}_{-0.027}$  & --&  $0.754^{+0.049}_{-0.041}$\\ 
$S_8$ & -- & -- & -- & -- & $0.739^{+0.055}_{-0.055} $ & -- & $0.711^{+0.080}_{-0.080}$ & --&$0.786^{+0.075}_{-0.072}$\\ 
$\gamma$ & -- & -- & --  & -- &  $0.527^{+0.096}_{-0.084}$ & $0.573^{+0.087}_{-0.101}$ &  $0.524^{+0.109}_{-0.086}$   & $0.640^{+0.042}_{-0.068}$   &$0.596^{+0.059}_{-0.063}$\\
\hline
&$f_2$CDM&&&&&&\\
$H_0$ &  $70.446^{+2.803}_{-2.779}$     & $68.341^{+7.113}_{-5.651}$  &$73.714^{+1.018}_{-1.007}$ &  $70.578^{+1.964}_{-1.952}$ & --&--&--&  $73.772^{+1.018}_{-1.011}$ &  $73.738^{+1.018}_{-1.013}$\\ 
$\Omega_m$ &$0.334^{+0.046}_{-0.060}$   &$0.318^{+0.057}_{-0.057}$  &   $0.368^{+0.023}_{-0.034}$  &$0.312^{+0.058}_{-0.052}$ & $0.301^{+0.065}_{-0.074}$   & $0.210^{+0.068}_{-0.056}$ & $0.217^{+0.064}_{-0.043}$ & $0.322^{+0.020}_{-0.019}$&$0.361^{+0.027}_{-0.035}$\\ 
$p$ & $0.837^{+0.110}_{-0.136}$ &  $0.803^{+0.108}_{-0.086}$  &  $0.938^{+0.045}_{-0.073}$& $0.822^{+0.098}_{-0.088}$  & $0.797^{+0.147}_{-0.346}$ & $0.721^{+0.174}_{-0.181}$ & $0.898^{+0.072}_{-0.113}$ &$0.962^{+0.028}_{-0.049}$&  $0.952^{+0.035}_{-0.058}$\\ 
$r_d$ & -- & $152.645^{+14.003}_{-14.383}$ &-- &   $147.520^{+3.541}_{-3.388}$& -- & -- & -- & --&-- \\ 
$M_{abs}$ & -- & -- & $-19.245^{+0.029}_{-0.029}$   &--   & -- & -- & -- & $-19.246^{+0.029}_{-0.030}$    &  $-19.245^{+0.029}_{-0.030}$ \\ 
$\zeta$ & $0.071^{+0.060}_{-0.047}$ & $0.075^{+0.040}_{-0.049}$ & $0.084^{+0.046}_{-0.049}$  &  $0.069^{+0.040}_{-0.044}$  & $0.154^{+0.097}_{-0.100}$ & $0.081^{+0.071}_{-0.063}$ & $0.074^{+0.077}_{-0.050}$ & $0.039^{+0.033}_{-0.026}$  & $0.090^{+0.046}_{-0.052}$ \\ 
$\sigma_8$ &-- & -- & -- & -- &  $0.698^{+0.063}_{-0.072}$ & --& $0.781^{+0.027}_{-0.026}$  & -- & $0.731^{+0.037}_{-0.029}$\\ 
$S_8$ &-- & -- & -- & -- & $0.699^{+0.078}_{-0.028}$& -- & $0.653^{+0.108}_{-0.108}$&-- & $0.801^{+0.050}_{-0.055}$\\
$\gamma$ & -- & -- & --&-- & $0.476^{+0.106}_{-0.057}$ &  $0.525^{+0.118}_{-0.091}$ &  $0.491^{+0.114}_{-0.067}$   & $0.668^{+0.023}_{-0.040}$   & $0.640^{+0.040}_{-0.056}$ \\ 
\hline
&{$f_3$CDM}&&&&&&\\
$H_0$ &  $71.132^{+2.407}_{-2.287}$ &  $69.116^{+7.277}_{-6.412}$  &  $73.824^{+1.027}_{-0.999}$& $70.413^{+1.829}_{-1.802}$  & --&--&-- & $73.926^{+1.003}_{-0.996}$&$73.815^{+1.015}_{-0.990}$ \\ 
$\Omega_m$ & $0.366^{+0.025}_{-0.041}$   & $0.365^{+0.025}_{-0.033}$ & $0.391^{+0.007}_{-0.013}$& $0.367^{+0.023}_{-0.032}$  & $0.309^{+0.059}_{-0.068}$& $0.192^{+0.051}_{-0.038}$ & $0.190^{+0.042}_{-0.027}$ & $0.362^{+0.016}_{-0.016}$& $0.389^{+0.008}_{-0.014}$ \\ 
$\Gamma$ & $3.247^{+1.204}_{-0.878}$   &$2.946^{+0.935}_{-0.667}$   &$2.360^{+0.468}_{-0.263}$   & | $2.870^{+0.867}_{-0.613}$  & $3.882^{+1.795}_{-1.324}$ & $3.754^{+1.629}_{-1.225}$& $2.969^{+1.129}_{-0.696}$ &  $2.192^{+0.286}_{-0.144}$  &  $2.323^{+0.437}_{-0.235}$\\ 
$r_d$ & -- &  $150.509^{+15.405}_{-14.283}$ & --&$147.335^{+3.499}_{-3.386}$ & -- & -- & -- &  --&-- \\ 
$M_{abs}$ &-- & --  & $-19.246^{+0.030}_{-0.029}$&-- & -- & -- & -- &  $-19.248^{+0.029}_{-0.029}$& $-19.247^{+0.029}_{-0.029}$  \\ 
$\zeta$ & $0.036^{+0.043}_{-0.026}$ & $0.041^{+0.019}_{-0.022}$& $0.018^{+0.022}_{-0.013}$  & $0.042^{+0.018}_{-0.022}$ &$0.142^{+0.102}_{-0.095}$ & $0.104^{+0.111}_{-0.074}$ & $0.049^{+0.073}_{-0.036}$ & $0.008^{+0.013}_{-0.006}$&$0.021^{+0.024}_{-0.015}$\\ 
$\sigma_8$ & -- &-- & --&-- & $0.675^{+0.044}_{-0.041}$ &  --   &  $0.765^{+0.024}_{-0.024}$ & --&  $0.700^{+0.016}_{-0.017}$  \\
$S_8$ & -- & -- & --&-- & $0.540^{+0.045}_{-0.042}$ & --& $0.603^{+0.042}_{-0.028}$ & --&$0.786^{+0.020}_{-0.023}$\\ 
$\gamma$ & -- & -- & --&-- &  $0.499^{+0.105}_{-0.070}$&  $0.471^{+0.082}_{-0.051}$ &$0.444^{+0.067}_{-0.032}$ & $0.686^{+0.011}_{-0.022}$ & $0.642^{+0.039}_{-0.054}$    \\ 
\hline
&$\Lambda$CDM&&&&&&\\
$H_0$ &$68.965^{+2.487}_{-2.287}$  & $69.210^{+7.222}_{-6.487}$ &  $73.587^{+0.998}_{-0.997}$ & $68.426^{+1.752}_{-1.760}$ & -- &--& -- &   $73.681^{+1.029}_{-1.010}$ &  $73.591^{+1.022}_{-1.011}$  \\ 
$\Omega_m$ &$0.348^{+0.036}_{-0.052}$  & $0.331^{+0.037}_{-0.029}$ & $0.365^{+0.024}_{-0.027}$&  $0.333^{+0.037}_{-0.029}$   & $0.298^{+0.066}_{-0.074}$ &  $0.230^{+0.050}_{-0.050}$   & $0.224^{-0.063}_{-0.045}$ &   $0.329^{+0.017}_{-0.017}$&  $0.362^{+0.025}_{-0.027}$\\  
$r_d$ & -- &  $145.698^{+15.006}_{-13.783}$& --&$147.199^{+3.521}_{-3.383}$ & -- & -- & --&-- &-- \\ 
$M_{abs}$ & -- & -- & $-19.244^{+0.029}_{-0.029}$ &--& -- & --& --&   $-19.245^{+0.030}_{-0.029}$&$-19.244^{+0.029}_{-0.030}$  \\ 
$\zeta$ & $0.049^{+0.052}_{-0.034}$& $0.028^{+0.025}_{-0.019}$ &  $0.058^{+0.042}_{-0.038}$ &$0.028^{+0.025}_{-0.020}$  & $0.135^{+0.106}_{-0.093}$ & $0.128^{+0.102}_{-0.084}$   & $0.060^{+0.076}_{-0.043}$&   $0.020^{+0.024}_{-0.014}$& $0.058^{+0.043}_{-0.039}$\\ 
$\sigma_8$ & -- & --& --&-- & --  $0.753^{+0.045}_{-0.041}$& -- & $0.779^{+0.025}_{-0.024}$ & --& $0.726^{+0.028}_{-0.026}$  \\ 
$S_8$ & -- & --& --& -- & $0.750^{+0.016}_{-0.017}$ & -- & $0.785^{+0.038}_{-0.036}$& --&$0.789^{+0.041}_{-0.041}$ \\ 
$\gamma$ & --& -- & -- & --& $0.548^{+0.098}_{-0.095}$ & $0.511^{+0.094}_{-0.076}$ & $0.496^{+0.109}_{-0.068}$ & $0.673^{+0.020}_{-0.035}$&$0.627^{+0.047}_{-0.057}$ \\
\hline
\end{tabular}
\caption{{{Best-fit parameters calculated through MCMC simulations in the $\zeta \neq 0$, for $\Lambda$CDM, $f_{1\rm CDM}$, $f_{2\rm CDM}$ and $f_{3\rm CDM}$ models. Note that $H_{0}$ is given in $\mathrm{km/s/Mpc}$.}}}
\label{tab:best_fit_values}
\end{table*}
Using the best-fit values of these parameters, we study the bulk viscosity effect on the growth rate of cosmic structures by introducing additional damping in the evolution of density perturbations \cite{zimdahl1996bulk} This influences the distribution and clustering of matter \cite{yang2019challenging} which provides an analysis of the contribution of $f(Q)$ gravity, both with and without the presence of bulk viscosity, to study the density contrast \(\delta(z)\). This helps us to understand the formation of large-scale structures such as galaxies and clusters. 
\\
\\
The following priors have been considered during the simulations: \(H_0  = [60.0-80.0]\, \text{in km/s/Mpc} \), \( \Omega_m = [0.01, 0.4] \), \( r_d = [100, 200] \), \( \zeta = [0.0, 0.3] \), \(M_{abs} = [-22, -15] \),  \( \gamma = [0.4, 0.7) \), and \( \sigma_8 = [0.5, 1.0] \), \( n = [0.0, 0.5] \), \( p = [0.0, 1.0] \), and \( \Gamma = [2.0, 10] \) for all models.  We use the datasets methodology as discussed in Section \ref{datameth} to constrain the observational parameters.  This process reduces uncertainties since accurate constraints improve the reliability of the cosmological conclusions and drive advances in both observational methods and theoretical physics. 
\\
\\
The calculated best-fit parameter values are presented in the compacted form in Tables \ref{tab:best_fit_valueswithout} and \ref{tab:best_fit_values} for the $\Lambda$CDM and $f(Q)$ gravity models for the cases without and with the effect of bulk viscosity, respectively. 
The resulting contour plots are presented, Figures \ref{fig:enterlcdm1} and \ref{fig:enterlcdm2} for $\Lambda$CDM and \ref{fig:enterpower} and \ref{fig:enterpower1} for the $f_{1\text{CDM}}$ models,  in the case of without and with the effect of bulk viscosity, which is consistent at $1\sigma$ and $2\sigma$ confidence levels. Similarly, we present the contour plots in \ref{app}, for in Figures \ref{fig:expon}  and \ref{fig:expon1} for the $f_{2\text{CDM}}$ model; and  \ref{fig:log} and \ref{fig:log1} for the $f_{3\text{CDM}}$ model.

Moreover,  by resorting to the constrained values of $\Omega_m$ and $\sigma_8$ from Tables \ref{tab:best_fit_valueswithout} and \ref{tab:best_fit_values}, we have also calculated the corresponding values of $S_8$ defined as
\begin{equation} 
S_8 = \sigma_8 \sqrt{\frac{\Omega_m}{0.3}}\;.
\end{equation}
This is a combined parameter used especially in weak lensing and galaxy clustering analyses, better suited than $\sigma_8$ when degeneracies exist between $\sigma_8$ and $\Omega_m$ \cite{adil2024s}. The calculated value of $S_8$ gives crucial insight into the relationship between the Universe's matter density and the strength of its clustering. It helps us understand matter fluctuations, enabling comparisons between different observational data and theoretical models. Discrepancies in its value of the late and early measurements can indicate challenges to existing cosmological theories and suggest new physics, such as the nature of dark energy or modifications to gravity. From the constraining of the cosmological parameters (Tables \ref{tab:best_fit_valueswithout} and \ref{tab:best_fit_values}), we can determine the following properties of each of the three models. 
\subsubsection{Parameter results for $f_{1\rm{CDM}}$}
From our first model's constrained parameters, we note a minimal difference with or without the bulk viscosity on all parameters with the only clear exception being $\Omega_{m}$, where $f_{1\rm{CDM}}$ with bulk viscosity favors a higher matter density on all observation datasets except for the $f+f\sigma_{8}$ combined dataset. This is noteworthy since direct `distance ladder' measurement comparisons, such as the observations used in the paper, tend to favor a significantly lower value compared to CMB measurements. Therefore, reducing the difference of $\Omega_m$ for these two methods without impacting the remaining cosmological parameters is vital. The only other difference that we found in the constrained parameters for $f_{1\rm{CDM}}$ with and without bulk viscosity is a slight lowering of the $S_{8}$ parameter. Keeping in mind that even though the Planck collaboration reported a $S_8 = 0.830\pm 0.007$, a lower value of $S_{8}$ ranging between $S_8 = 0.759^{+0.025}_{-0.023}$  and  $0.772^{+0.018}_{-0.017}$, has been reported more recently by the DESb Y1 \& Y3 collaboration \cite{abbott2018dark, amon2022dark} respectively, matching within error our $f_{1\rm{CDM}}$ model with bulk viscosity. Furthermore, since $S_8$ relates to the degree of clustering of matter, this would indicate that adding bulk viscosity in $f_{1\rm{CDM}}$ leads to a universe with more matter but is slightly less clustered than in a non-viscous $f_{1\rm{CDM}}$ universe. This model is broadly explored in the work \cite{sahlu2025structure} without the viscous effect.  

\subsubsection{Parameter results for $f_{2\rm{CDM}}$}
For the case of $f_{2\rm CDM}$, the trend for a higher $\Omega_{m}$ value when bulk viscosity is included continues, but some variation in the constrained results occurs specifically in datasets that include $f$ where it seems to slightly favor a lower predicted $\Omega_m$ value. 

More interestingly, though, we obtain even lower values of $S_8 = 0.653^{+0.108}_{-0.108}$ and $\gamma = 0.476^{+0.106}_{-0.057}$ in the presence of the bulk viscosity effect. This can provide us with insights into the known $S_8$ tension. As for this result's interpretation, we can once again conclude that the addition of bulk viscosity in a $f_{2\rm CDM}$ based universe will increase its matter content, but even less clustering of the matter will occur compared to a $f_{1\rm CDM}$ universe. 

\subsubsection{Parameter results for $f_{3\rm{CDM}}$}
In terms of constraining the cosmological parameters $f_{3\rm{CDM}}$, the variations in the predicted best-fit values of $\Omega_m$ are now clearly seen when bulk viscosity is included. These values range from as high as $0.391^{+0.007}_{-0.013}$ (significantly higher than the CMB reported values) on the SNIa \textit{PantheonP + SH0ES} dataset, to as low as $0.190^{+0.042}_{-0.027}$ on the combined $\textit{f} + f\sigma_8$ dataset. That said, this wide range obtained is not purely due to the contribution of bulk viscosity, seeing that the $f_{3\rm{CDM}}$ model already obtained a wide range of $\Omega_m$ results; bulk viscosity only enhances the increase or decrease (seen in $f_{2\rm{CDM}}$'s $f$-related results). Continuing on the previous note trends, we obtain even lower $S_8$ results when bulk viscosity is included, and at this moment, these values seem too low based on our current best-fit $\Lambda$CDM results. Without bulk viscosity $f_{3\rm{CDM}}$ is still in an acceptable range, but is noticeably lower than expected. 
\subsubsection{General parameter remarks}

Interestingly, for the $\Lambda$CDM model, the variation in $\Omega_{m}$ is also noticeable but not as pronounced as seen with the $f_{3\rm{CDM}}$ model's results. The addition of bulk viscosity also increases the variation seen here. Therefore, this might be an indication that the $\Omega_{m}$ tension is more related to the observations themselves, rather than the models fitted onto the observations. Further testing would be necessary. Furthermore, the $H_{0}\, (\text{km/s/Mpc})$ tension between the BAO/corrected covariance matrix CC and \textit{PantheonP + SH0ES} is clearly evident in all our results, particularly for the $\Lambda$CDM and $f1CDM$ models. Consequently, these models do not contribute to resolving the $H_{0}$ tension. 
\begin{figure}[h!]
     \includegraphics[scale=0.37]{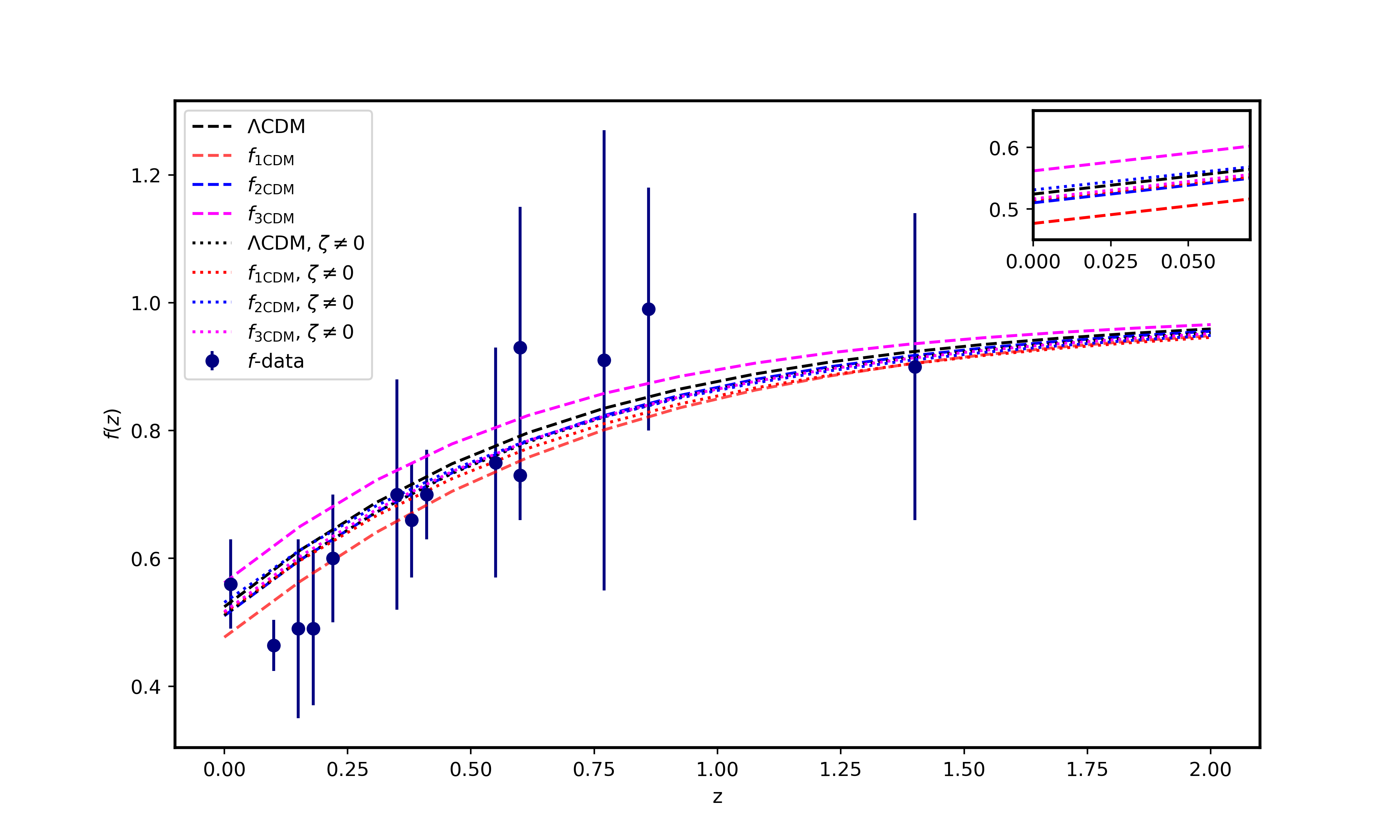}
   \caption{Evolution of the growth-rate  \(f(z)\) versus redshift for both cases: without bulk viscosity \((\zeta= 0)\) and with bulk viscosity $\zeta = 0.058,\, 0.077,\, 0.090,\, 0.021$ for $\Lambda$CDM, $f_{1\rm CDM}$, $f_{2\rm CDM}$ and $f_{3\rm CDM}$ models, respectively,
   as per the $PantheonP\,+\,SHOES\,+\,f\sigma_8$ best fits found on Table \ref{tab:best_fit_values}. The top right panel represents the zoomed-in view of the growth rate in the range of redshift, $z$ = 0 to 0.06, indicating the minimal effects of the bulk viscosity.
   }
    \label{fig:enterfplotfz}
 \end{figure}
 %
 \begin{figure}[h!]
     \includegraphics[scale=0.37]{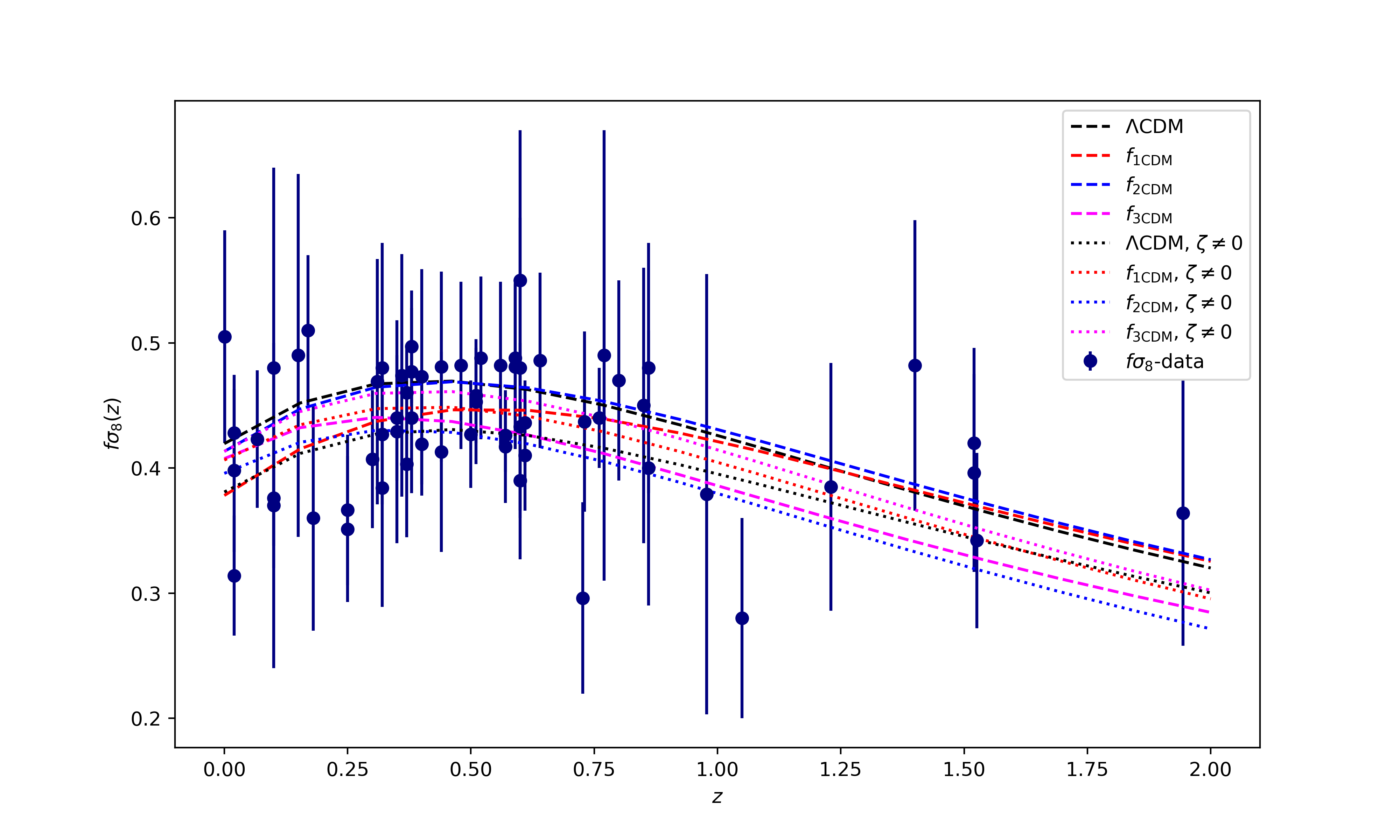}
   \caption{Evolution of the redshift space distortion  \(f\sigma_8(z)\) versus redshift for both cases: without bulk viscosity 
   $(\zeta=0)$  and with bulk viscosity 
   $\zeta = 0.058,\, 0.077,\, 0.090,\, 0.021$ for $\Lambda$CDM, $f_{1\rm CDM}$, $f_{2\rm CDM}$ and $f_{3\rm CDM}$ models, respectively, as per the $PantheonP\,+\,SHOES\,+\,f\sigma_8$   
   best fits found on Table \ref{tab:best_fit_values}.}
    \label{fig:enterf8}
\end{figure}
We also provide numerical results of the growth factor \(f(z)\) as a function of cosmological redshift, without and with bulk viscosity effects. These results are displayed in Figure \ref{fig:enterfplotfz}, is shows the results without and with bulk viscosity for the three different $f(Q)$ gravity models, including $\Lambda$CDM. Analyzing these results allows us to measure how density perturbations evolve, eventually leading to the formation of cosmic structures such as galaxies and clusters of galaxies within the framework of the $f(Q)$ gravity model, both with and without the influence of bulk viscosity. From this plot, we observe that the $f(Q)$ gravity models without the bulk viscosity effect align more closely with the predictions of the $\Lambda$CDM model, particularly at higher redshift.
We also notice the slight deviation at lower redshift, see Figure  \ref{fig:enterfplotfz}.  In the same manner, the redshift-space distortion \(f\sigma_8(z)\) as a function of the cosmological redshift is shown both without and with the effects of bulk viscosity in Figure \ref{fig:enterf8}, for all models. From this diagram, we note that the theoretical predictions of all three $f(Q)$ gravity models align well with observational data from redshift-space distortions (\textit{f}$\sigma_8$), compared to the $\Lambda$CDM model, therefore visually increasing the validity of these models. We have taken the required best-fit value of the parameters from Tables \ref{tab:best_fit_valueswithout}, and \ref{tab:best_fit_values} for the case of {\textit{PantheonP + SH0ES} + $f\sigma_8$}.

\subsection{Statistical analysis}
The statistically determined values for both models ($\Lambda$CDM and $f(Q)$ gravity models) using {CC}, \textit{BAO}, {\textit{PantheonP + SH0ES}}, \textit{f} and $f\sigma_8$ and their combined datasets: {\textit{CC}+ \textit{BAO}},  {\textit{PantheonP + SH0ES} + \textit{f}} and  {\textit{PantheonP + SH0ES} + $f\sigma_8$} datasets have also been calculated. These include likelihood (${\mathcal{L}(\hat{\theta}|data)}$),  chi-square ($\chi ^{2}$), reduced chi-square ($\chi^{2}_{\nu}$),  Akaike Information Criterion (AIC) \cite{Akaike1974}, change of AIC ($|\Delta\rm{AIC}|$), Bayesian Information Criterion (BIC) \cite{Schwarz1978}, and  change of BIC ($|\Delta\rm{BIC}|$). Since these statistical values are essential techniques in model selection, the goodness-of-fit compares the viability of the $f(Q)$ gravity models with that of $\Lambda$CDM. 
We consider the $\Lambda$CDM as the ``accepted'' model to justify the $f(Q)$ gravity model based on the AIC and BIC criteria for comparison purposes. These criteria allow us to establish the acceptance or rejection of a particular $f(Q)$ gravity model. The AIC and BIC values are calculated using the following equations:
$\rm{AIC} = \chi ^{2} +2\kappa,\quad \rm{and}\quad\rm{BIC} = \chi ^{2} +\kappa\log(N_i),$ where $\chi^{2}$ is calculated using the model's likelihood function $\mathcal{L}(\hat{\theta} |data)$ value and $\kappa$ is the number of free parameters for the particular model. At the same time, $N_i$ is the number of data points for the $i^{th}$ dataset. We apply the Jeffreys scale \cite{nesseris2013jeffreys} range to quantify whether a particular $f(Q)$ gravity model should be ``accepted'' or ``rejected'' relative to $\Lambda$CDM. Accordingly, $\Delta IC \leq 2$ means that for the fitted data the proposed theoretical model holds \textrm{substantial observational support}, $ 4 \leq \Delta IC \leq 7$  means  \textrm{less observational support}, and finally $\Delta IC \geq 10$ means \textrm{no observational support}.\footnote{\textbf{N.B} To avoid any confusion,  $\Delta IC $ represents both $\Delta\rm{AIC} $ and  $\Delta\rm{BIC} $. }

The statistics-based information has been provided in Tables \ref{stastical1} and \ref{stastical2} for both with and without the bulk viscosity effects. Note that the discussions regarding the constraining statistical analysis for this work take into account the CMB based \cite{aghanim2020planck} results: We assume the base $\Lambda$CDM cosmology and the inferred (model-dependent) late-Universe parameter values for the Hubble constant \(H_{0} = 67.4\pm 0.5\) km/s/Mpc; matter density parameter \(\Omega_m= 0.315\pm0.007\); and matter fluctuation amplitude \(\sigma_8 = 0.811\pm 0.006.\)
%
\begin{table*}[h!]
\caption{ Values of ${\mathcal{L}(\hat{\theta}|data)}$, $\chi^2$, $\chi^{2}_\nu$, AIC, $|\Delta\rm{AIC}|$, BIC, and $|\Delta\rm{BIC}|$ for both the $\Lambda$CDM model and the $f(Q)$ gravity models using {CC}, \textit{BAO},  {\textit{PantheonP + SH0ES}}, $f$, $f\sigma_8$,  $f$+ $f\sigma_8$, {\textit{CC}+ \textit{BAO}},  { \textit{PantheonP + SH0ES} + \textit{f}} and  {\textit{PantheonP + SH0ES} +$f\sigma_8$} datasets for the case of $\zeta= 0$.}
\label{stastical1}
    \begin{tabular}{|lrrrrrrrr|}
        \hline
        \textbf{Data } &  & \textbf{$\mathcal{L}(\hat{\theta}|data)$} & \textbf{$\chi ^{2}$} &$\chi^{2}_\nu$ & AIC &\textbf{$|\Delta\rm{AIC}|$} & BIC & \textbf{$|\Delta\rm{BIC}|$} \\
        \hline
        \textbf{$\Lambda{\rm CDM}$}& &&&&&&&\\ 
        & &&&&&&&\\
        {CC}   & &  -7.259& 14.519  &  0.500 & 18.520 & -- & 21.388 & --\\
        {BAO}   & &-6.378 & 12.756 & 1.417&18.756 & -- & 20.211 & --\\
        {\textit{PantheonP + SH0ES}} &&  -726.322 &    1452.645&    0.878  &  1458.646 & -- &  1474.884 & --\\
       {\textit{CC}+ \textit{BAO}} &&-13.723&  27.447 & 0.686 & 33.447& -- & 38.731 & --\\
        \textit{$f$} &  &-3.082& 6.164 &  0.513& 10.164& -- & 11.442& --\\
        {$f\sigma_8$} & &-15.990& 31.9808 & 0.507  & 37.981& --&44.550& --\\
        {$f+f\sigma_8$} & &-19.365& 38.730 & 0.503 & 44.731& --& 51.877& --\\
        {\textit{PantheonP + SH0ES}  + $f$} & & -729.775&  1459.551 &  0.875 &  1467.552 &--&  1489.237 & --\\
         {\textit{PantheonP + SH0ES} + $f\sigma_8$} & & -742.462&   1484.925 & 0.864&  1494.925 &--& 1522.184& --\\
        \hline
        \textbf{$f_{\rm 1CDM}$}& &&&&&&&\\ 
        & &&&&&&&\\
        \textit{CC}   & &  -7.318& 14.636  & 0.522  & 20.636 & 2.115 & 24.938 & 3.549\\
        \textit{BAO}   & &-6.896 & 13.793 & 1.724&21.794 & 3.038& 23.733 & 3.522\\
        {\textit{PantheonP + SH0ES}} &&  -726.220&  1452.440 &  0.878   & 1460.440 & 1.794 & 1482.091 &    7.207\\
       {\textit{CC}+ \textit{BAO}} && -14.169   &     28.338  &   0.726  & 36.339 & 2.891 &    43.383 &  4.653
 \\
        {$f$} &  &-3.398& 6.796&  0.617 & 12.797& 2.633 & 14.714& 3.272\\
        {$f\sigma_8$} & &-15.874& 31.749 & 0.5121 & 39.750& 1.769&48.508& 3.958\\
        {$f+f\sigma_8$} & &-19.399& 38.799 & 0.510 & 46.799& 2.068& 56.328& 4.451\\
        {\textit{PantheonP + SH0ES}  + $f$} & &  -729.603 &  1459.206 &  0.875  & 1469.207 &  1.655&  1496.312 &7.076\\
    {\textit{PantheonP + SH0ES}  + $f\sigma_8$} & &  -742.087&  1484.174 &  0.864  &  1496.175 &1.250& 1528.886 &  6.701\\
    \hline
     \textbf{$f_{\rm 2CDM}$}& &&&&&&&\\ 
        & &&&&&&&\\
        {CC}   & & -7.3017& 14.603  & 0.521  & 20.603 & 2.083 &24.905  & 3.668\\
        {BAO}   & &-7.864 & 15.728    & 1.962& 23.729 & 4.973& 25.668 & 5.457\\
        {\textit{PantheonP + SH0ES}} &&   -727.731 & 1455.462 &     0.880 &  1463.463 & 4.817 & 1485.114 & 10.230
\\
       {\textit{CC}+ \textit{BAO}} && -15.001 &    30.002     &    0.769   &  38.002 & 4.555 & 45.047 &6.316
\\
        {$f$} &  &-2.743 & 5.487 &  0.498 & 11.487 & 1.323  &  13.404& 1.962 \\
        {$f\sigma_8$} & &-17.697& 35.394& 0.570   & 43.395& 5.414&52.153& 7.614\\
        {$f+f\sigma_8$} & &-19.625 & 39.250 & 0.516  & 47.250&  2.519&  56.778& 4.902\\
        {\textit{PantheonP + SH0ES}  + $f$} & &   -731.209 &  1462.418 & 0.877 & 1472.418 & 4.866& 1499.524& 10.287\\
     {\textit{PantheonP + SH0ES} + $f\sigma_8$} & &  -743.738&  1487.476 &  0.866 &1499.476&4.551 & 1532.187 & 10.003
 \\
        \hline
     \textbf{$f_{\rm 3CDM}$}& &&&&&&&\\ 
        & &&&&&&&\\
        \textit{CC}   & &-7.377 & 14.7539  & 0.5269  & 20.754 & 2.234   &25.056  & 3.668\\
        \textit{BAO}   & & -7.454 & 14.908   & 1.863 &22.909 & 4.152& 24.848 & 4.637\\
        {\textit{PantheonP + SH0ES}} && -727.034 &1454.068   &   0.879 & 1462.069 & 3.423    &1483.720 & 8.836\\
       {\textit{CC}+ \textit{BAO}} &&-15.033 & 30.066     &   0.770 & 38.067 & 4.620& 45.112 & 6.381\\
        {$f$} &  &-2.7182  & 5.436 &  0.4942  & 11.436 &  1.272 &  13.354& 1.911\\
        {$f\sigma_8$} & &-16.495& 32.989 & 0.532   & 40.990& 3.009&49.749 & 5.199 \\
        {$f+f\sigma_8$} & &-19.781 & 39.562 & 0.520   & 47.563& 2.832&  57.091& 5.214\\
        {\textit{PantheonP + SH0ES}  + $f$} & &  -734.908 &  1469.816  &  0.882  &  1479.817 &  12.265& 1506.923 &  17.686\\
     {\textit{PantheonP + SH0ES} + $f\sigma_8$} & & -744.131&   1488.263 &   0.866 &  1500.264 &5.339&  1532.974 &   10.790
 \\
        \hline
    \end{tabular}
\end{table*}
\begin{table*}[h!]
\caption{Values of ${\mathcal{L}(\hat{\theta}|data)}$, $\chi^2$, $\chi^{2}_\nu$, AIC, $|\Delta\mathrm{AIC}|$, BIC, and $|\Delta\mathrm{BIC}|$ for both the  $\Lambda$CDM model and the $f(Q)$ gravity models using \textit{CC}, \textit{BAO}, \textit{PantheonP + SH0ES}, $f$, $f\sigma_8$, $f + f\sigma_8$, \textit{CC + BAO}, \textit{PantheonP + SH0ES + f}, and \textit{PantheonP + SH0ES + $f\sigma_8$} datasets for the case of $\zeta \neq 0$.}
\label{stastical2}
    \begin{tabular}{|lrrrrrrrr|}
        \hline
        \textbf{Data } &  & \textbf{$\mathcal{L}(\hat{\theta}|data)$} & \textbf{$\chi ^{2}$} &$\chi^{2}_\nu$ & AIC &\textbf{$|\Delta\rm{AIC}|$} & BIC & \textbf{$|\Delta\rm{BIC}|$} \\
        \hline
        \textbf{$\Lambda{\rm CDM}$}& &&&&&&&\\ 
        & &&&&&&&\\
        \textit{CC}   & & -7.418& 14.837   &   0.529   &  20.837 & 2.318 & 25.139 & 3.752\\
        \textit{BAO}   & & -6.640 & 13.281 & 1.660& 21.282 & 2.525 & 23.221 & 3.010\\
        {\textit{PantheonP + SH0ES}} &&  -726.312 &   1452.624 &   0.878  &   1460.624 & 1.979  & 1482.275 & 7.391\\
       {\textit{CC}+ \textit{BAO}} && -13.971& 27.943  &  0.716& 35.94&  2.496  &  42.988  &  4.257\\
        \textit{f} &  &-2.785& 5.571& 0.506& 11.572& 1.407 & 13.489& 2.047\\
        {$f\sigma_8$} & &-15.984 & 31.969 & 0.515  & 39.970 & 1.989&48.728& 4.179\\
        {$f+f\sigma_8$} & &-19.490&  38.981& 0.512 &  46.982& 2.251&56.5107&4.633  \\
        {\textit{PantheonP + SH0ES} + \textit{f}} & &  -730.028& 1460.056 &  0.876 &  1470.0561 & 2.504&   1497.162  & 7.926\\
        {\textit{PantheonP + SH0ES} + $f\sigma_8$} & & -742.355&  1484.710 &  0.864& 1496.710  & 1.785&  1529.421&   7.237 \\
        \hline
        \textbf{$f_{1{\rm CDM}}$}& &&&&&&&\\ 
        & &&&&&&&\\
        \textit{CC}   & & -7.501& 15.003  &   0.555 &   23.004 & 4.484  &  28.740 & 7.352 \\
        \textit{BAO}   & & -7.326 & 14.652 &2.0932& 24.652 & 5.896  & 27.077 & 6.866\\
        {\textit{PantheonP + SH0ES}} &&    -726.558 &  1453.116&   0.879 &  1463.116 &  4.471  & 1490.180 &15.296
\\
       {\textit{CC}+ \textit{BAO}} &&  -14.553&  29.107&  0.766 & 39.108 &5.660  &  47.914  & 9.183
\\
        \textit{f} &  &-2.834& 5.668 &  0.566 & 13.669& 3.504  & 16.225&  4.783\\
        {$f\sigma_8$} & &-15.943 & 31.887 & 0.522  & 41.887 & 3.906&52.835& 8.286 \\
        {$f+f\sigma_8$} & &-19.357&  38.714& 0.516 &  48.714& 3.983 &60.624& 8.747 \\
        {\textit{PantheonP + SH0ES}  + \textit{f}} & &    -729.688&  1459.376&0.876 &1471.377  &3.825&  1503.904 &  14.667\\
        {\textit{PantheonP + SH0ES}  + $f\sigma_8$} & & -742.391&  1484.783 &   0.865 &  1498.783 & 3.858&  1536.946 &  14.762
\\
        \hline
         \textbf{$f_{2{\rm CDM}}$}& &&&&&&&\\ 
        & &&&&&&&\\
        \textit{CC}   & & -7.2881& 14.576   &   0.539 &   22.576 & 4.056  &  28.312 &  6.924\\
        \textit{BAO}   & & -8.069 & 16.138 &2.305& 26.138 &7.382 & 28.563 & 8.352  \\
        {\textit{PantheonP + SH0ES}} &&   -727.226 &    1454.452& 0.880 &   1464.452 &   5.807  &   1491.516& 16.632 \\
       {\textit{CC}+ \textit{BAO}} &&  -15.248 &  30.497 & 0.802 & 40.498 &7.050 &  49.304 & 10.573
\\
        \textit{f} &  &-2.710&  5.421 &  0.542  & 13.421& 3.257  & 15.977& 4.535\\
        {$f\sigma_8$} & &-16.551 & 33.103 & 0.542  &43.103 & 5.122  &54.052& 9.502\\
        {$f+f\sigma_8$} & &-19.437&   38.874& 0.518 &  48.874&  4.143 &60.784&   8.908 \\
        {\textit{PantheonP + SH0ES} +  \textit{f}} & &   -730.650&     1461.300 & 0.877  &  1473.301& 5.749& 1505.828&     16.591\\
        {\textit{PantheonP + SH0ES}  + $f\sigma_8$} & & -743.065&   1486.130 &  0.866&1500.130 &  5.205 & 1538.293 &  16.109\\
        \hline
       { \textbf{$f_{3{\rm CDM}}$}}& &&&&&&&\\ 
        & &&&&&&&\\
        \textit{CC}   & & -7.531& 15.061   &  0.557  &   23.062   & 4.542  & 28.798 & 7.410 \\
        \textit{BAO}   & & -7.400  & 14.800 &2.114 & 24.801 &6.044 & 27.225   & 7.014\\
        {\textit{PantheonP + SH0ES}} &&  -727.172&  1454.344 &   0.880&  1464.345  &  5.699 &  1491.4081& 16.525\\
       {\textit{CC}+ \textit{BAO}} &&  -14.562&  29.125 & 0.766  & 39.126&  5.678 &  47.932 &  9.201
  \\
        \textit{f} &  &-2.817 & 5.634 &   0.563   & 13.63&  3.470  & 16.191 & 4.748 \\
        {$f\sigma_8$} & &-16.210 & 32.421 & 0.531 &42.421& 4.440 &53.369& 8.820  \\
        {$f+f\sigma_8$} & &-20.023&   40.046& 0.534 &  50.047& 5.316 &61.957& 10.080\\
        {\textit{PantheonP + SH0ES} + \textit{f}} & & -735.435&   1470.871 & 0.883  &  1482.871 & 15.320&  1515.398  & 26.162\\
        {\textit{PantheonP + SH0ES} +  $f\sigma_8$} & & -744.146 &  1488.292 &  0.8673& 1502.293 & 7.367 &1540.455  & 18.271 \\
        \hline
    \end{tabular}
\end{table*}

\subsubsection{Statistical analysis results for $f_{1\rm{CDM}}$}
From a statistical analysis perspective, for the $f_{1\rm{CDM}}$ model without bulk viscosity, no observational dataset outright rejects the model. In fact, the highest $\Delta$IC value is $7.207$ (which falls just outside the ``less observational support'' category) for the \textit{PantheonP + SH0ES} dataset, a significant step forward in establishing the model’s validity as a viable alternative. Moreover, we obtained four datasets with $\Delta$AIC scores that firmly fall within the ``substantial observational support'' category relative to the $\Lambda$CDM model, while the remaining datasets still yield $\Delta$AIC values above the ``less support'' threshold. Although no $\Delta$BIC values reached the substantial support category, all values  (except two that narrowly miss this category) still fall at least within the ``less observational support or higher'' range. These findings suggest that there is at least moderate observational support for the $f_{1\rm{CDM}}$ model as an alternative to the $\Lambda$CDM model.

In contrast, when bulk viscosity is incorporated, three $\Delta$BIC values — specifically for datasets that include the \textit{PantheonP + SH0ES} data — outright reject the model as a viable alternative. This outcome arises from the introduction of the extra free parameter $\zeta$, which does not improve the $\chi^{2}$ value and, in many cases, leads to a poorer fit. For example, for the CC dataset, the $\Lambda$CDM model yields a $\chi^{2}$ value of $14.519$, whereas the $f_{1\rm{CDM}}$ model without bulk viscosity (with one additional free parameter relative to $\Lambda$CDM) obtains $14.636$, and with bulk viscosity (with two extra free parameters) the value increases to $15.003$.

Therefore, statistically, the addition of bulk viscosity to the $f_{1\rm{CDM}}$ model does not improve the fit; rather, it degrades the model’s performance. This shifts the model from being a reliable alternative with at least moderate observational support to one that, according to the $\Delta$BIC, is largely unsupported, with only two $\Delta$BIC value falling in the ``less observational support'' category. Nonetheless, from the $\Delta$AIC perspective, the model still retains some degree of observational support and is not entirely rejected, although none of the $\Delta$AIC values reach the substantial category as observed in the case without bulk viscosity.

\subsubsection{Statistical analysis results for $f_{2\rm{CDM}}$}
From a statistical analysis perspective, the $f_{2\rm{CDM}}$ model without bulk viscosity appears less consistent with the data than the $f_{1\rm{CDM}}$ model, as indicated by its AIC and BIC values. Specifically, while the $f$ dataset yielded values in the ``substantial observational support'' category, three datasets (those that include the \textit{PantheonP + SH0ES} data) obtained $\Delta$BIC values that outright reject the model. 

This discrepancy highlights an important issue: since the \textit{PantheonP + SH0ES} dataset is by far the largest and most complete dataset in our analysis, its restrictive nature demands a very close fit from any viable model. Statistically, larger datasets tend to have smaller uncertainties, resulting in more precise parameter estimates and a more sharply peaked likelihood function. Consequently, even minor discrepancies between the model predictions and the observed data can yield significantly higher $\chi^{2}$ values, thereby affecting the AIC and BIC scores. In contrast, smaller datasets, which typically have larger uncertainties, allow for more flexibility in the fit. Therefore, a model’s inability to adequately capture the trends in the \textit{PantheonP + SH0ES} data is particularly detrimental to it case for being a viable alternative model relative to the $\Lambda$CDM model.

When bulk viscosity is incorporated into the $f_{2\rm{CDM}}$ model, the overall effect is similar to those seen in the $f_{1\rm{CDM}}$ scenario: the additional free parameter $\zeta$ increases the  $\Delta$AIC and $\Delta$BIC values, thereby penalizing the model without improving the fit. In this case, no dataset yields $\Delta$AIC or $\Delta$BIC values within the substantial observational support category. In summary, the $f_{2\rm{CDM}}$ model — both with and without bulk viscosity — performs worse than both versions of the $f_{1\rm{CDM}}$ model.

\subsubsection{Statistical analysis results for $f_{3\rm{CDM}}$}
From a statistical analysis perspective, the $f_{3\rm{CDM}}$ model results, both with and without bulk viscosity, are broadly similar to those obtained for the $f_{2\rm{CDM}}$ model. The main difference is a slight $\Delta$AIC or $\Delta$BIC improvement for the \textit{PantheonP + SH0ES} data and now not being outright rejected; however, this improvement is not statistically significant. Moreover, on average, $f_{3\rm{CDM}}$ obtained too high $\Omega_{m}$ and too low $\sigma_8$ values, especially when bulk viscosity is included, indicating that this model is not a viable alternative model.

\subsubsection{General statistical analysis remark}
In summary, the inclusion of bulk viscosity does not improve the likelihood function of the models sufficiently to justify its additional free parameters, as evidenced by the consistently higher $\Delta$AIC and $\Delta$BIC values—even in the viscous $\Lambda$CDM case. Nevertheless, the introduction of bulk viscosity has provided valuable insights for future $f(Q)$ gravity models. Our analysis revealed that in some observational datasets, bulk viscosity exacerbates the increase in $\Omega_m$. This effect may help to narrow the discrepancy between local distance ladder measurements and CMB estimates. Conversely, in other $f$-related datasets, the inclusion of bulk viscosity appears to worsen the $\Omega_m$ tension. A similar behavior is observed in the viscous $\Lambda$CDM model, suggesting that these discrepancies may depend, at least in part, on the specific characteristics of the datasets utilized.

Among the three $f(Q)$ models analyzed, only the $f_{1\rm{CDM}}$ model exhibits stable parameter values, with $\Omega_{m}$ and $\sigma_{8}$ remaining within the expected error margins. Notably, $f_{1\rm{CDM}}$ model without bulk viscosity does not lead to any outright observational rejections and even achieves four cases of “substantial observational support.” Although the addition of bulk viscosity slightly degrades the statistical performance of $f_{1\rm{CDM}}$ (with rejections appearing only in the $\Delta$BIC criterion), the model remains the most viable alternative overall. While AIC and BIC are standard tools for model selection, and our analysis is consistent with their use, exploring alternative model selection criteria in future work may provide additional insights into these trends.

\section{Conclusions}\label{conculusions}
This work extensively analyzed the late-time cosmic accelerated expansion and the evolution and formation of large-scale structures through the utilization of recent observational cosmological data. In particular, we studied these phenomena within the framework of $f(Q)$ gravity models, by probing the impact of the bulk viscosity. This study aimed to gauge the validity and broader implications of the $f(Q)$ gravity models, with a focus placed on the role of bulk viscosity in $f(Q)$ gravity. 
\\
\\
The theoretical foundation of $f(Q)$ gravity was expanded in Section \ref{theory}, wherein bulk viscosity was integrated into the effective pressure of the matter fluid, denoted as $\bar{p} = p-3\zeta_1 H$. In Section \ref{models} we introduced three distinct classes of $f(Q)$ gravity models, namely: (i) power-law, (ii) exponential, and (iii) logarithmic models (termed $f_{1\rm{CDM}}$, $f_{2\rm{CDM}}$, and $f_{3\rm{CDM}}$,  respectively) to assess their viability in different cosmological scenarios. The derived normalized Hubble parameter expressions $E(z)$ for each model are presented in Eqs. \eqref{power}, \eqref{expo} and \eqref{log} and the distance modulus expressions \(\mu(z)\) in Eq. \eqref{distancemodules1} were instrumental in evaluating the validity of these $f (Q)$-gravity models along with the $\Lambda$CDM predictions when confronted with recent cosmic measurements. 
\\
\\
To investigate the growth of structure in a viscous fluid, the theoretical framework of density contrast \(\delta(z)\) was detailed in Section \ref{perturbations} within the 1+3 covariant formalism. This section presents the evolution equations for scalar perturbations, incorporating spatial gradients of matter, volume expansion, and gauge-invariant variables defining the fluctuations of the energy density and momentum terms of the non-metricity treated as an extra ``fluid'', as outlined in Eq. \eqref{35} and \eqref{oooo}. Implementing scalar- and harmonic decomposition techniques, the second-order differential equation for the density contrast is derived in Eq. \eqref{densitycontrast123}  for all three models. Additionally, the growth factor \(D(z)\), growth rate \(f(z)\), and \(f \sigma_8(z)\) are formulated in the context of the $f(Q)$ gravity model to understand the formation and evolution of large-scale structures in the Universe. These quantities are instrumental in testing the viability of the $f(Q)$ gravity theory and constraining parameters of cosmological models through observational data, which provide crucial insights into the Universe's evolution from early epochs to late time.
\\
\\
{In Section \ref{resultdiscussion}, extensive analysis has been conducted on constraining cosmological parameters utilizing \texttt{Kosmulator's} MCMC simulations on various observational datasets. To accomplish this, firstly, we constrained the parameters \(\Omega_m\),  $H_0\,\mathrm{(km/s/Mpc)}$,  \(r_d\), \(M_{abs}\), \(\gamma\), \(\sigma_8\) and exponents \(n\), \(p\) and \(\Gamma\) including the bulk viscosity coefficient \(\zeta\) values as depicted in the compacted form in Figures \ref{fig:enterlcdm1} and \ref{fig:enterlcdm2} for $\mathrm{\Lambda}$CDM and Figures \ref{fig:enterpower} and \ref{fig:enterpower1} for $f_{1\rm{CDM}}$ at $1\sigma$ and $2\sigma$ confidence levels. In the same manner, the contour plots for the $f_{2\rm{CDM}}$ and $f_{3\rm{CDM}}$ models are also found in the \ref{app}, see Figures \ref{fig:expon} - \ref{fig:log1}. 

The above best-fit cosmological parameter values for each cosmological model are provided in Tables \ref{tab:best_fit_valueswithout} and \ref{tab:best_fit_values} for both $\Lambda$CDM and $f(Q)$ gravity models without and with the effect of bulk viscosity. From these tables, we noticed that the bulk viscosity has a minimal effect on all models for all parameters as supported by the growth rate diagram presented in Figure \ref{fig:enterfplotfz}. The only exception to this is for the parameters $\Omega_m$ increasing and $\sigma_8$ decreasing in the presents of bulk viscosity, especially noticeable on the $f_{3}\text{CDM}$ model; however, there seems to be some indication that this tension might be related in part to the $f$ and $f\sigma_8$ dataset itself rather than it being related to the addition of bulk viscosity, which just seems to enhance the effect already present in the $\Lambda$CDM case. This needs further testing. The diagram of the growth rate and redshift space distortion fitting has been presented in Figures \ref{fig:enterfplotfz}  and \ref{fig:enterf8} to see the significant effect of $f(Q)$ gravity models for the structure formation.
\\
\\
{The corresponding detailed statistical analysis was presented in Tables \ref{stastical1} and \ref{stastical2} to quantify the viability of the $f(Q)$ gravity models for the cases of without and with bulk viscosity effects, respectively.  From these constrained parameter values and statistical results, we noted that for all measurements, $f_{1\rm{CDM}}$ without bulk viscous resulted in the most plausible alternative model relative to the $\Lambda$CDM model by obtaining the lowest $|\Delta\rm{BIC}|$, and concluded that $f_{1\rm CDM}$ is the only model that is suitable with and without viscous effect among the three $f(Q)$ gravity models for all datasets. {The combined datasets, \textit{ PantheonP + SH0ES +f} and \textit{ PantheonP + SH0ES + $f\sigma_8$}  resulted  $\Delta\rm BIC \geq 10$ (models being rejected) for  $f_{1\rm{CDM}}$ (with viscous effect), $f_{2\rm{CDM}}$ and $f_{3\rm{CDM}}$ (for both cases: with and without viscous effect) based on our statistical criteria. }
\\
\\{Thus, we conclude that, by resorting to the observational datasets considered in this work, the statistical acceptability of one of the $f(Q)$ models, namely the power-law $f_{1 \rm CDM}$ model without viscous fluid medium is a good candidate model.  Some additional tests, such as the study of CMB physics, may put more stringent constraints on the cosmological viability of $f(Q)$ models when endowed with viscosity.}

\section*{Data Availability}
We have used the publicly available cosmological probes, as listed in
Section \ref{datameth}.
\section*{Acknowledgments}
The authors thank Adri\'an Casado-Turri\'on and Francisco J. Maldonado-Torralba for helpful discussions.
AdlCD acknowledges support from BG20/00236 action (MCINU, Spain), NRF Grant CSUR23042798041, CSIC Grant COOPB23096, Project SA097P24 funded by Junta
de Castilla y Le\'on (Spain) and Grant PID2021-122938NB-I00 funded by MCIN/AEI/10.13039/ 501100011033 and by {\it ERDF A way of making Europe}. The authors would like to thank Prof. J. van der Walt for computational time on his server. 
\providecommand{\newblock}{}

\appendix
\section{Appendix}\label{app}
In the following, the MCMC simulations contour plots are shown for models $f_{2\rm CDM}$  and $f_{3\rm CDM}$, both with and without bulk viscosity.  
\begin{figure*}[h!]
 		\centering
 		\includegraphics[scale=0.20]{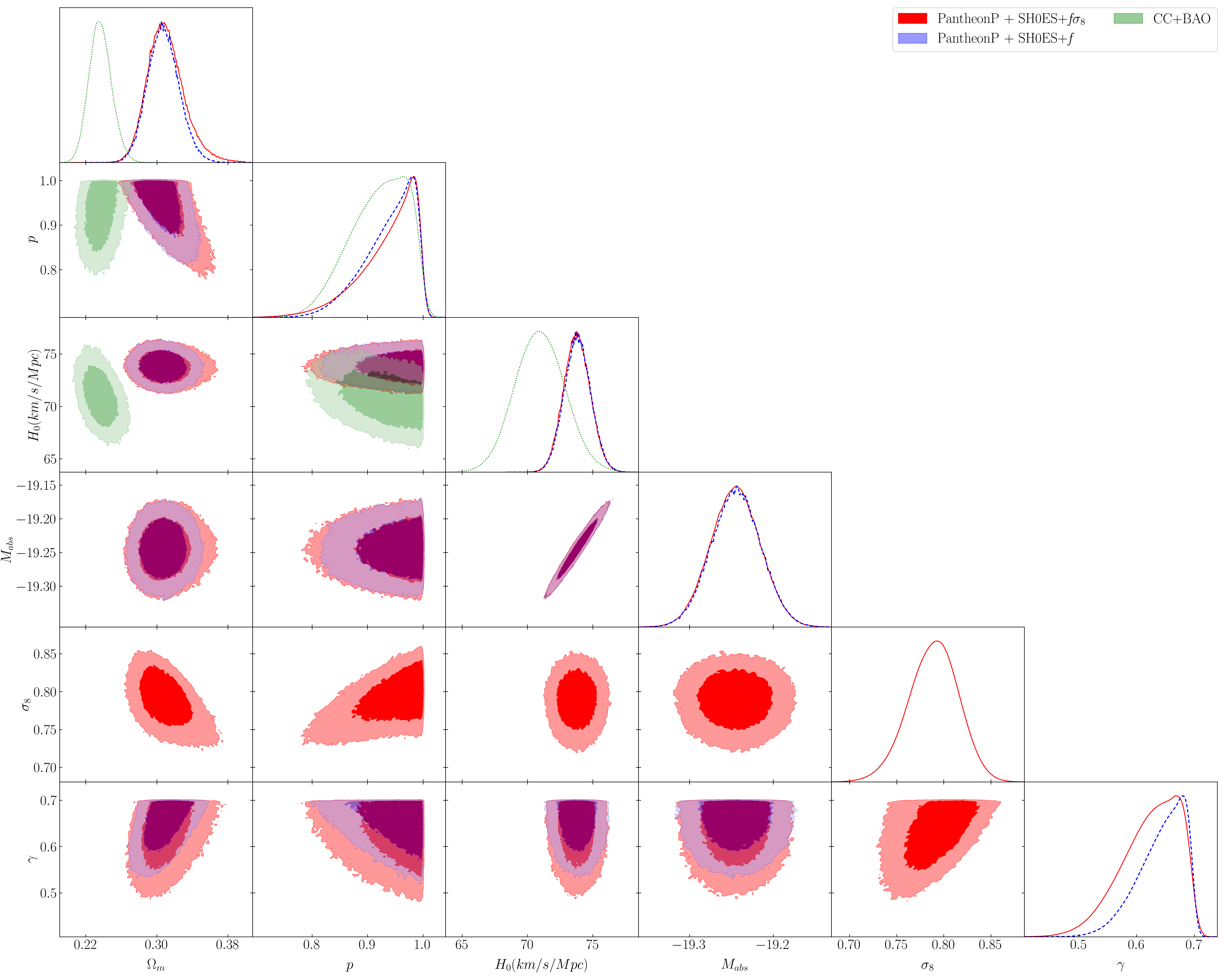}
   \caption{Best-fit parameter values for $f_{1\rm CDM}$ in the case of ($\zeta = 0$), using the datasets {\textit{PantheonP + SH0ES} + $f\sigma_8$}, { \textit{PantheonP + SH0ES} +$f$}, and {CC+BAO}, evaluated at $1\sigma$ and $2\sigma$ confidence levels.}
    \label{fig:expon}   
\end{figure*}
\begin{figure*}[h!]
 		\centering
 		\includegraphics[scale=0.20]{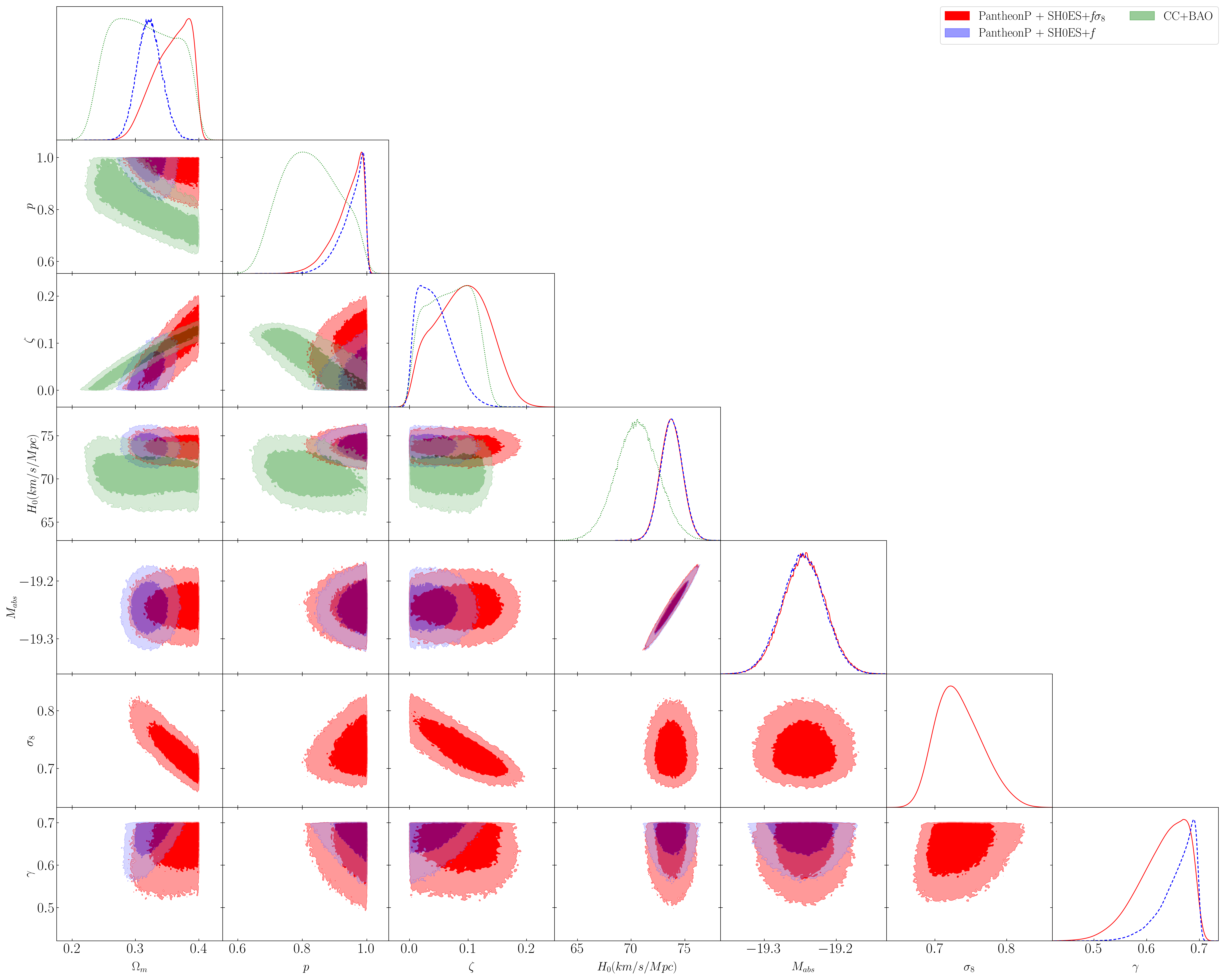}
   \caption{Best-fit parameter values for $f_{1\rm CDM}$ in the case of ($\zeta \neq 0$), using the datasets {\textit{PantheonP + SH0ES} + $f\sigma_8$}, { \textit{PantheonP + SH0ES} +$f$}, and {CC+BAO}, evaluated at $1\sigma$ and $2\sigma$ confidence levels.}
    \label{fig:expon1}   
\end{figure*}
\begin{figure*}[h!]
 		\centering
 		\includegraphics[scale=0.20]{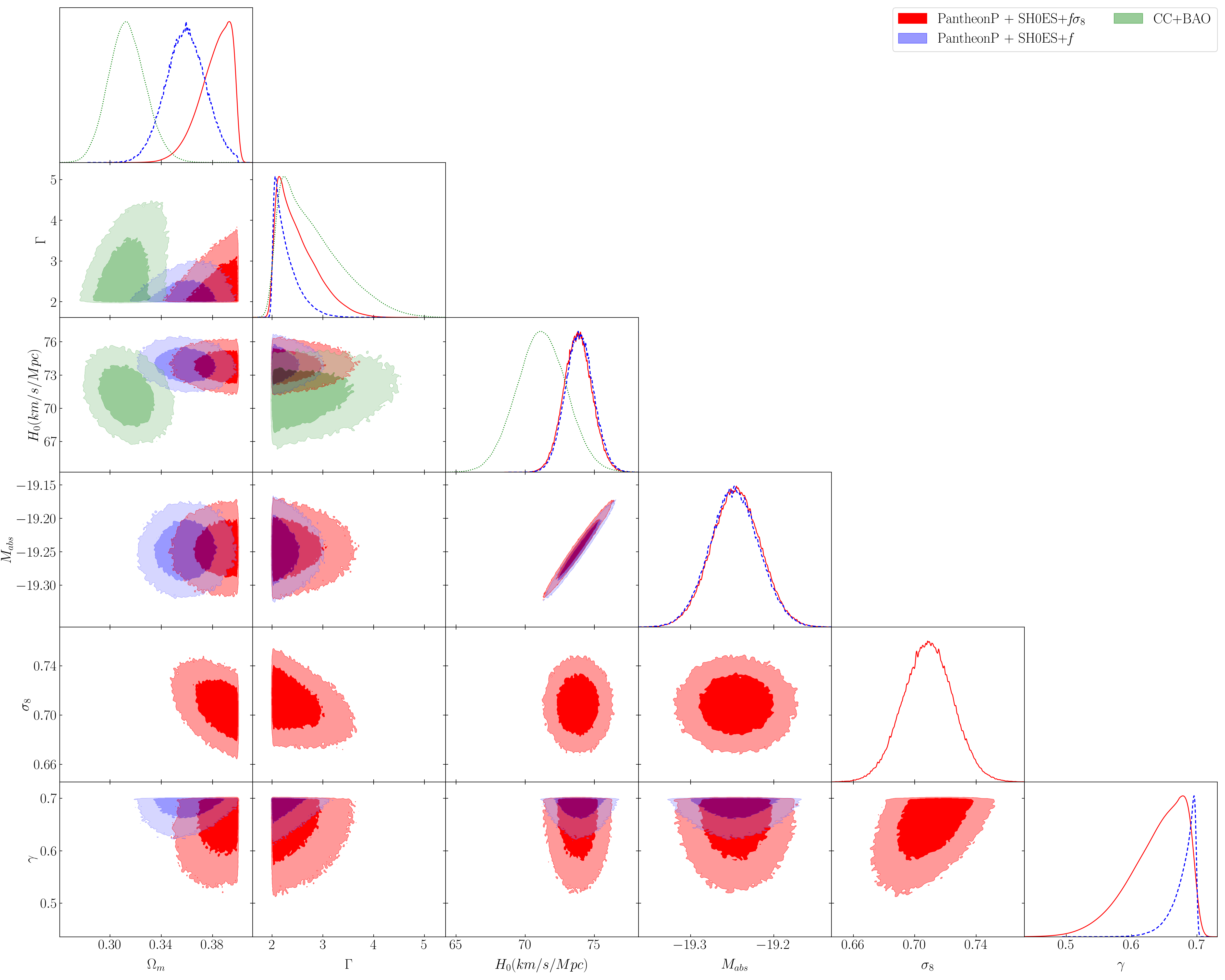}
   \caption{Best-fit parameter values for $f_{3\rm CDM}$ in the case of ($\zeta = 0$), using the datasets {\textit{PantheonP + SH0ES} + $f\sigma_8$}, { \textit{PantheonP + SH0ES} +$f$}, and {CC+BAO}, evaluated at $1\sigma$ and $2\sigma$ confidence levels.}
    \label{fig:log}   
\end{figure*}
\begin{figure*}[h!]
 		\centering
 		\includegraphics[scale=0.20]{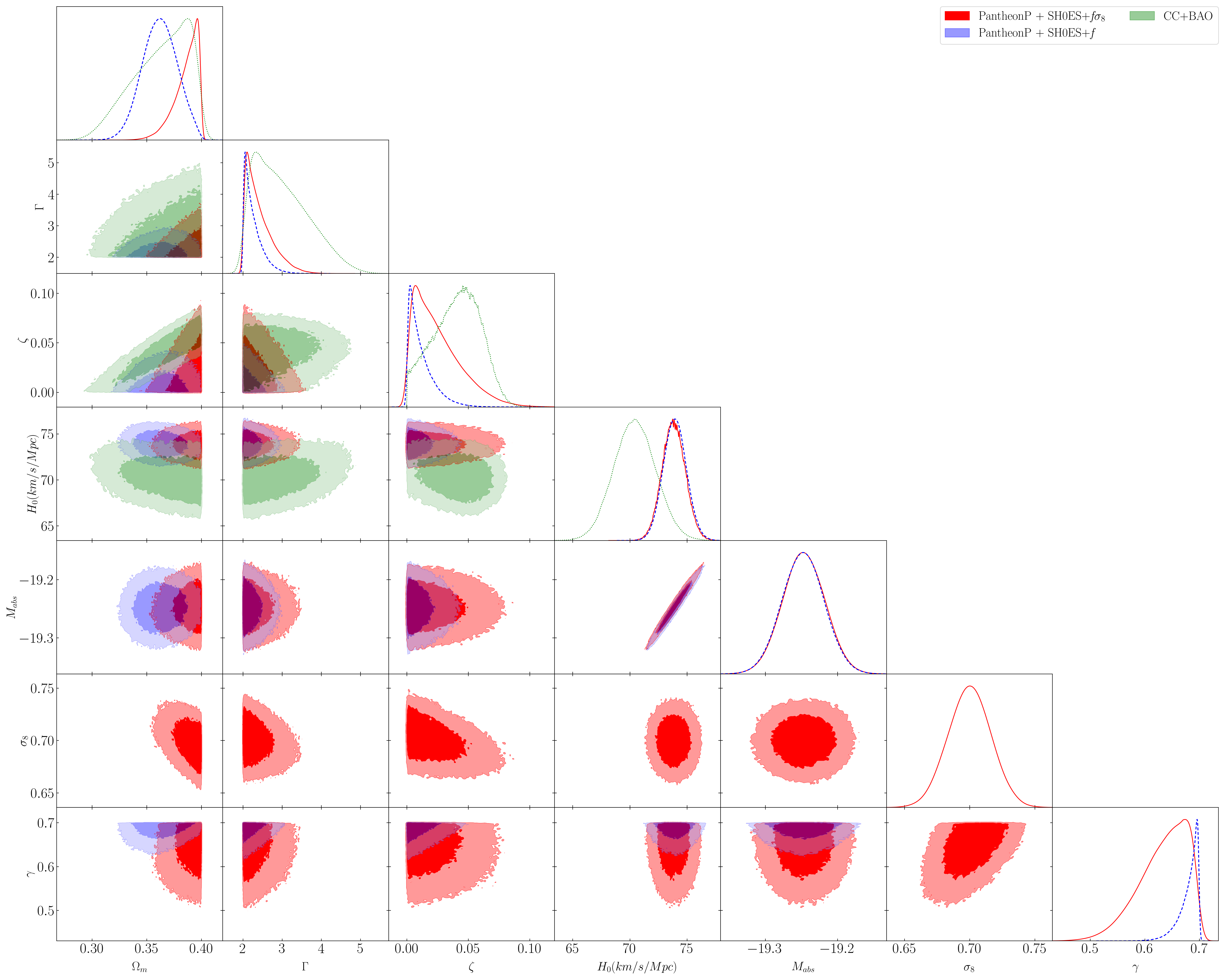}
   \caption{Best-fit parameter values for $f_{3\rm CDM}$ in the case of ($\zeta \neq 0$), using the datasets {\textit{PantheonP + SH0ES} + $f\sigma_8$}, { \textit{PantheonP + SH0ES} +$f$}, and CC+BAO, evaluated at $1\sigma$ and $2\sigma$ confidence levels.}
    \label{fig:log1}   
\end{figure*}
\end{document}